\documentclass[prb,twocolumn,floatfix,showpacs,superscriptaddress]{revtex4-1}
\usepackage{graphics}
\usepackage{epsfig}
\usepackage{times}
\usepackage{bm}
\usepackage{braket}
\usepackage{color}
\usepackage{amssymb}

\graphicspath{{./figures/}}

\usepackage{amsmath}	

\begin{document}
\title{Critical spin-$\frac{1}{2}$ tetramer compound CuInVO$_5$: Exploring the vicinity\\ of two multimerized singlet states}
\author{Sahinur Reja}
\affiliation{School of Mathematics and Physics, The University of Queensland, Brisbane, Queensland 4072, Australia}
\author{Satoshi Nishimoto}
\affiliation{Department of Physics, Technical University Dresden, 01069 Dresden, Germany}
\affiliation{Institute for Theoretical Solid State Physics, IFW Dresden, 01069 Dresden, Germany}

\date{\today}

\begin{abstract}
Using the density-matrix renormalization group technique, we study a one-dimensional spin-$\frac{1}{2}$ Heisenberg chain consisting of coupled tetramers as an effective spin model for copper vanadate CuInVO$_5$. We obtain the ground-state phase diagram as a function of intra-tetramer and inter-tetramer exchange interactions, exhibiting two multimerized singlet phases : one is characterized by the formation of tetramer-singlet units; the other by the formation of dimer-singlet pairs. We show that the finite spin gaps in both the singlet phases smoothly vanish at the phase boundary: a second order phase transition defining a quantum critical point (QCP). The phase boundary is also captured by the fact that the central charge is unity at the phase boundary and zero otherwise in the thermodynamic limit. It is interesting that the dimer-singlet state is interpreted as a Haldane state with the hidden $\mathbb{Z}_2 \times \mathbb{Z}_2$ symmetry breaking. We further demonstrate that the experimental magnetization curve (which starts increasing with zero or tiny field) can be reasonably explained only by assuming the exchange parameters of CuInVO$_5$ to be very close to the phase boundary. Thus, we argue that CuInVO$_5$ may be a first example material which at ambient pressure stands near a QCP between two singlet phases. By varying the balance of exchange interactions with pressure, a transition from N\'eel to either of the singlet phases could be observed.
\end{abstract}

\maketitle

\section{Introduction}

The concept of quantum criticality is now widely believed to be central to understand the physics of strongly correlated system. At zero temperature, a second- order quantum phase transition is associated with a quantum critical point (QCP)~\cite{Coleman05,Sachdev11} where the critical fluctuations are scale-invariant and the system belongs to a universality class characterized by critical exponents, independent of the microscopic details of the system~\cite{Moriya85,Binney92,Timusk99}. In addition, the physical properties over a wide range of temperatures above a QCP could be influenced by the critical fluctuations. Actually, a variety of exotic (non-Fermi-liquid) behaviors due to strong quantum fluctuations, i.e., quantum critical phenomena, have been reported in various systems such as high-$T_{\rm c}$ superconductivity~\cite{Keimer15}, heavy fermions~\cite{Lohneysen07,Gegenwart08}, and iron pnictites~\cite{Shibauchi14}, etc. So, quantum critical phenomena are not just an object of theoretical interest but the key to explain experimental observations.

In recent years, the significant progress of ultracold atomic scinece has led to a number of experimental studies to find the quantum criticality by controlling the interaction parameters in optical lattices~\cite{Zhang12}. Quantum magnets can also provide a fertile playground to study the critical phenomena. In these materials, the quantum phase transitions could occur through a QCP by controlling the exchange interactions by the application of external field~\cite{Coldea10} and/or of pressure~\cite{Paglione03}. A well-known example is thallium copper chloride TlCuCl$_3$~\cite{Oosawa03,Ruegg03,Tanaka03} having effective spin-$\frac{1}{2}$ Cu$^{2+}$ ions. At ambient pressure, all the spins pair into spin-singlet dimers and the system is in a gapped antiferromagnet. With increasing pressure, the system goes into a N{\'e}el state through a QCP located at $P_{\rm c}=1.07$ kbar~\cite{Ruegg04}. Near the QCP, the emergence of a well-defined longitudinal mode in the spin excitations was also reported as a signature of semi-classical N\'eel order~\cite{Ruegg08}. 

\begin{figure}
  \includegraphics[width=0.95\columnwidth]{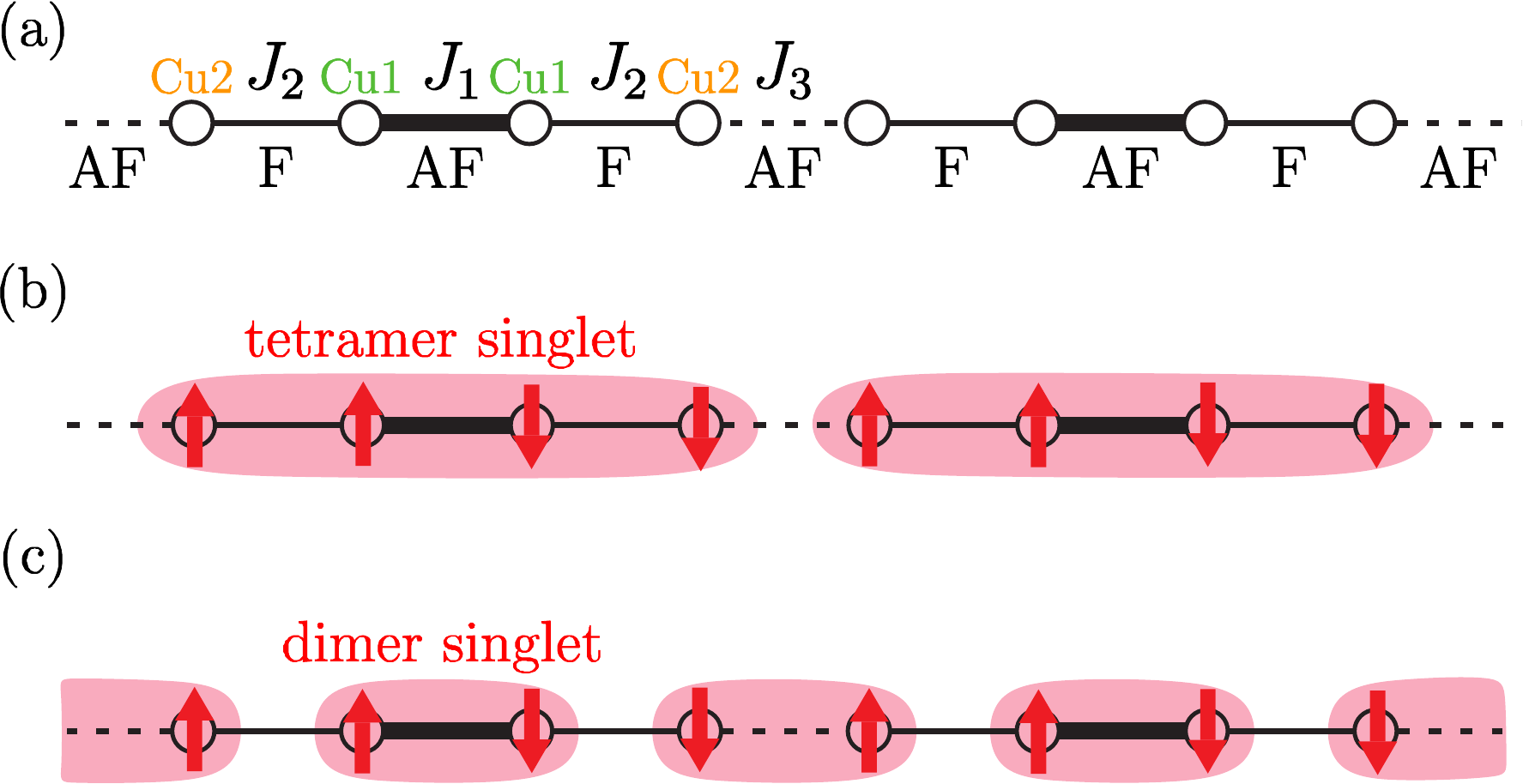}
  \caption{(a) Lattice structure of the one-dimensional coupled-tetramer Heisenberg model. Schematic pictures of (b) tetramer-singlet and (c) dimer-singlet states. Red ellipse denotes a spin-singlet formation.
  }
  \label{lattice}
\end{figure}

In this Letter, we consider copper vanadate CuInVO$_5$ as another candidate material near the QCP~\cite{Hase16,Singhania18}. Crystallographically, there are two different Cu sites having spin-$\frac{1}{2}$ on each Cu$^{2+}$ ion~\cite{Moser99}. A tetramer, formed with two Cu1 and two Cu2 sites, is the unit cell of a possible one-dimensional (1D) spin model as presented in Fig.~\ref{lattice}(a). We call this model as ``tetramer chain'' hereafter. From the crystal structure, the interactions between tetramers seem to be relatively weak. Based on the fittings of experimental data within an isolated spin-$\frac{1}{2}$ tetramer calculation, the effective exchange interactions have been estimated as $J_1=240 \pm 20$ K, $J_2=-142 \pm 10$ K, and $J_3= 30 \pm 4$ K~\cite{Hase16}. A certain magnitude of interchain coupling must also exist since a N\'eel order below $T_{\rm N} = 2.7$K has been observed~\cite{Hase16}. 

The tetramer chain contains two multimerized singlet phases in the ground state; depending on the exchange parameters, tetramer-singlet [Fig.~\ref{lattice}(b)] or dimer-singlet [Fig.~\ref{lattice}(c)] state appears (see below for details). Of particular interest is that the exchange parameters for this compound are most likely in a competing region of the two singlet states~\cite{Singhania18}; and the system should be highly sensitive to external influences like magnetic field, pressure as well as temperature. Nowadays, various other tetramer compounds exist: Cu$_2$CdB$_2$O$_6$ (spin-$\frac{1}{2}$)~\cite{Carmalt95,Hase15}, SeCuO$_3$ (spin-$\frac{1}{2}$)~\cite{Effenberger86,Zivkovic12}, Cu$_2$Fe$_2$Ge$_4$O$_{13}$ (spin-$\frac{1}{2} \&$  spin-$\frac{5}{2}$)~\cite{Matsumoto10,Masuda04}, and Rb$_2$Ni$_2$Mo$_3$O$_{12}$ (spin-$1$)~\cite{Klevtsova82,Hase17}. Further discoveries of exciting physics in these compounds would be reasonably expected. Nevertheless, there has been few theoretical study on the tetramer compounds.

Motivated by the above, we study the Heisenberg tetramer chain using the density-matrix renormalization group (DMRG) method~\cite{White92} to investigate the ground-state properties of CuInVO$_5$. We determine the ground-state phase diagram as a function of intra-tetramer ferromagnetic (FM) and inter-tetramer antiferromagnetic (AFM) exchange interactions, containing the two multimerized singlet phases, i.e., tetramer-singlet and dimer-singlet phases. We show that at the phase boundary the spin gaps in the both singlet states continuously approach to zero and the central charge is unity. We thus confirm that the phase transition is of the second order, i.e., identified as a QCP. By analyzing the experimental magnetization, we find that the possible parameters of CuInVO$_5$ are nearly on the QCP at ambient pressure. The observed N\'eel order may be realized if the (nearly) critical tetramer chains are coupled~\cite{Yasuda05}. By applying pressure and/or by lowering temperature, a transition from the N\'eel to singlet phases could be observed. The tetramer compound CuInVO$_5$ would be a first material near the QCP between two singlet phases. Therefore, this paper provides a deeper insight into the quantum criticality.

The paper is organized as follows: In Sec.~II our Hamiltonian of the tetramer chain is explained and the applied numerical method is described. In Sec.~III we present our numerical results. The ground-state phase diagram is determined base on the results of spin gap and central charge. The possibility of hidden $Z_2 \times Z_2$ symmetry breaking order is also discussed. Furthermore, by analyzing the experimental magnetization curve, it is confirmed that the tetramer compound CuInVO$_5$ stands really close to a QCP. In Sec. IV we give a conclusion and discussion/speculation on the tetramer compound CuInVO$_5$.

\section{Model and method}

The lattice structure of tetramer chain is presented in Fig.~\ref{lattice}(a). The model Hamiltonian reads
\begin{align}
\nonumber
\mathcal{H}=\sum_{i=1}^{N_{\rm t}}[&J_1\vec{S}_{i,2}\cdot\vec{S}_{i,3}+J_2(\vec{S}_{i,1}\cdot\vec{S}_{i,2}+\vec{S}_{i,3}\cdot\vec{S}_{i,4})\\
+&J_3\vec{S}_{i,4}\cdot\vec{S}_{i+1,1}+h\sum_{j=1}^4S^z_{i,j}]
\label{ham}
\end{align}
where $\vec{S}_{i,j}$ is the spin-$\frac{1}{2}$ operator at $j$-th site in $i$-th tetramer, $N_{\rm t}$ is the total number of tetramers, $J_1$ and $J_2$ are intra-tetramer couplings, $J_3$ is inter-tetramer coupling, and $h$ is external field. The total length of system is $L=4N_{\rm t}$. According to Ref.\onlinecite{Hase16}, we assume $J_1(>0)$, $J_2(<0)$, and $J_3(>0)$ to be AFM, FM, and AFM, respectively. 

One can easily imagine the ground state in two limiting cases: (I) For $J_1 >|J_2| \gg J_3$, each tetramer is in a singlet state as sown in Fig.~\ref{lattice}(b), whose wavefunction is approximately
$|{\rm g.s.}\rangle_{\rm tet}=\prod_{i=1}^{N_{\rm t}}|T_i\rangle$
with a singlet formation in tetramer,
$|T_i\rangle=\frac{1}{2}(|\uparrow\rangle_{i,1}|\downarrow\rangle_{i,4}-|\downarrow\rangle_{i,1}|\uparrow\rangle_{i,4})(|\uparrow\rangle_{i,2}|\downarrow\rangle_{i,3}-|\downarrow\rangle_{i,2}|\uparrow\rangle_{i,3})$,
where $|\uparrow\rangle_{i,j}$ and $|\downarrow\rangle_{i,j}$ denote spin states at $j$-th site in $i$-th tetramer. This singlet state is exact in the limit $J_1 \gg |J_2| \gg J_3$. Although the other terms are gradually mixed at finite $|J_2|/J_1$, $|{\rm g.s.}\rangle_{\rm tet}$ is still a good approximation for the ground state at $J_1 \gtrsim |J_2|$. Thus, we call this state ``tetramer-singlet state''. The spin gap is $\Delta=\frac{1}{2}(J_2+\sqrt{J_1^2-2J_1J_2+4J_2^2}-\sqrt{J_1^2+J_2^2})$. (II) For $J1, J_3 \gg |J_2|$, each of $J_1$ and $J_3$ bonds forms a dimer singlet pair as sown in Fig.~\ref{lattice}(c). The wavefunction is a product state of the singlet pairs, namely,
$|{\rm g.s.}\rangle_{\rm dim}=\prod_{i=1}^{N_{\rm t}}|D_i\rangle$
with
$|D_i\rangle=\frac{1}{2}(|\uparrow\rangle_{i,2}|\downarrow\rangle_{i,3}-|\downarrow\rangle_{i,2}|\uparrow\rangle_{i,3})
(|\uparrow\rangle_{i,4}|\downarrow\rangle_{i+1,1}-|\downarrow\rangle_{i,4}|\uparrow\rangle_{i+1,1})$.
We call this state ``dimer-singlet state''. The spin gap is $\Delta=\min(J_1,J_3)$.

We use the DMRG method to study the ground-state properties of system \eqref{ham}. The periodic boundary conditions are applied. Since the entanglement is relatively short ranged in the ground state, we can perform accurate calculations for chains with up to $L=200$ by keeping up to $3000$ density-matrix eigenstates in the renormalization procedure. For three-leg ladder with $L\times3=64\times3$, we keep up to $5000$ density-matrix eigenstates.

\begin{figure}
  \includegraphics[width=0.9\columnwidth]{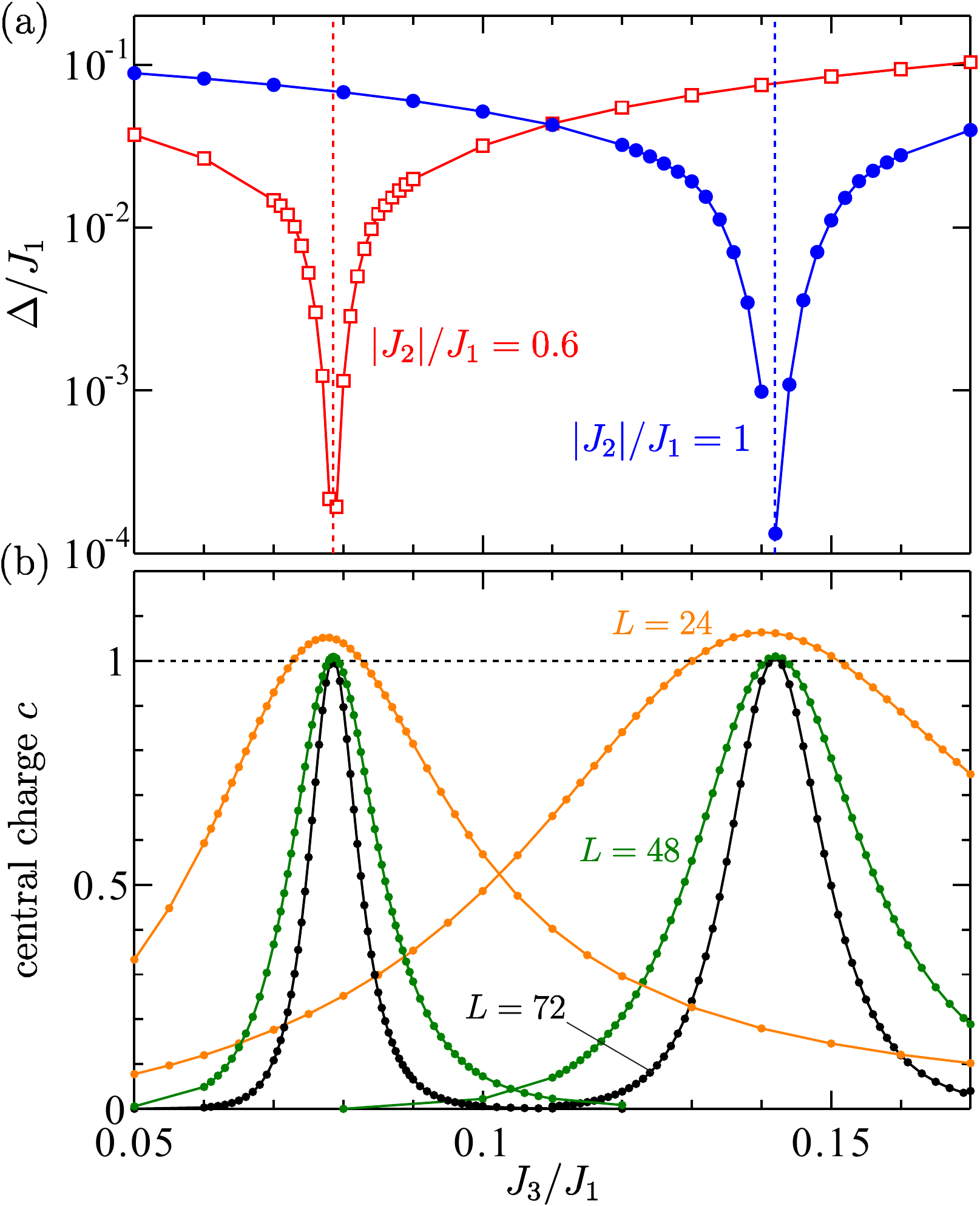}
  \caption{(a) Spin gap as a function of $J_3/J_1$ for $|J_2|/J_1=0.6$ and $1$. The dotted line denotes the boundary between tetramer-singlet and dimer-singlet phases. (b) System-size dependence of the central charge as a function of $J_3/J_1$ for $|J_2|/J_1=0.6$ and $1$.
  }
  \label{gap_c}
\end{figure}

\section{Results}

\subsection{Spin gap}

Both the singlet states are characterized by finite spin gap. The spin gap is defined as
\begin{align}
\Delta=\lim_{L\to\infty}E_0(L,1)-E_0(L,0),
\label{spingap}
\end{align}
where $E_0(L,S^z_{\rm tot})$ is the total ground-state energy of the system with length $L$ and the $z$-component of total spin $S^z_{\rm tot}$. In Fig.~\ref{gap_c}(a), the spin gap is plotted as a function of $J_3/J_1$ for $|J_2|/J_1=0.6$ and $1$ on semilogarithmic scale. For both the $|J_2|/J_1$ values, the spin gap drops continuously to zero at a single $J_3/J_1$ value. This may suggest a direct second-order transition between tetramer-singlet and dimer-singlet phases; namely, the system is gapless only at the phase boundary. The spin gap increases with the distance from the phase boundary. In the limit $J_3 \gg |J_2|$, it saturates to $\Delta/J_1=1$ being the energy to break a singlet pair on $J_1$ bond; whereas in the limit $J_3 \ll |J_2|$, it saturates at $\Delta/J_1=1/4$ because the system \eqref{ham} is equivalent to isolated spin-$1$ $J_1$ dimers. Furthermore, it is interesting that the dimer-singlet state is interpreted as a Haldane state~\cite{Affleck87}. As shown below, by examining the string order parameter~\cite{deNijs89,Agrapidis18}, we have confirmed that the hidden $\mathbb{Z}_2 \times \mathbb{Z}_2$ symmetry breaking~\cite{Oshikawa92} occurs in the whole region of dimer-singlet phase.

\subsection{Central charge}

The central charge $c$ provides definitive information on the universality class of $(1+1)$ dimensional system~\cite{Cardy96}. A system in the Tomonaga-Luttinger phase belongs to the Gaussian universality class ($c=1$) and $c<1$ is expected for the gapped phase from the renormalization in the massive region. The central charge can be numerically calculated through the von Neumann entanglement entropy $S_L(l)=-{\rm Tr}_l \rho_l \log \rho_l$, where $\rho_l={\rm Tr}_{L-l}\rho$ is the reduced density matrix of the subsystem with length $l$ and $\rho$ is the full density matrix of the whole system with length $L$. Using the conformal field theory (CFT), the relation between $S_L(l)$ and $c$ has been derived~\cite{Affleck91,Holzhey94,Calabrese04}: $S_L(l)=\frac{c}{3}\ln\left[\frac{L}{\pi}\sin\left(\frac{\pi l}{L}\right)\right]+s_1$, where $s_1$ is a non-universal constant. A prime objective of using this formula is to estimate the central charge~\cite{Laflorencie06,Legeza07}. For the system \eqref{ham} the central charge is obtained via~\cite{cc}
\begin{align}
c=\frac{3\left[S_L\left(\frac{L}{2}-8\right)-S_L\left(\frac{L}{2}\right)\right]}
{\ln\left[\cos\left(\frac{8\pi}{L}\right)\right]}.
\label{eq:centralcharge}
\end{align}
Note that the difference of $S_L(l)$ is taken between $l=L/2$ and $L/2-8$ since the unit cell contains four sites and periodic boundary conditions are applied. More details are explained in Appendix B.

In Fig.~\ref{gap_c}(b) the central charge is plotted as a function of $J_3/J_1$ for $|J_2|/J_1=0.6$ and $1$ for several system lengths $L$. In each case a single Lorentzian-like peak is obtained. It appears that the peak becomes sharper but keeps its height around $c\approx1$ with increasing $L$; in fact, it is extrapolated to a $\delta$-peak of height $c=1$ in the thermodynamic limit. The finite-size scaling analyses of the width, height, and position are given in Appendix B. This $\delta$-peak clearly indicates a gapless point corresponding to the QCP between the two singlet phases. The critical $J_3/J_1$ values agree well with that estimated from the spin gap.

\begin{figure}
  \includegraphics[width=0.9\columnwidth]{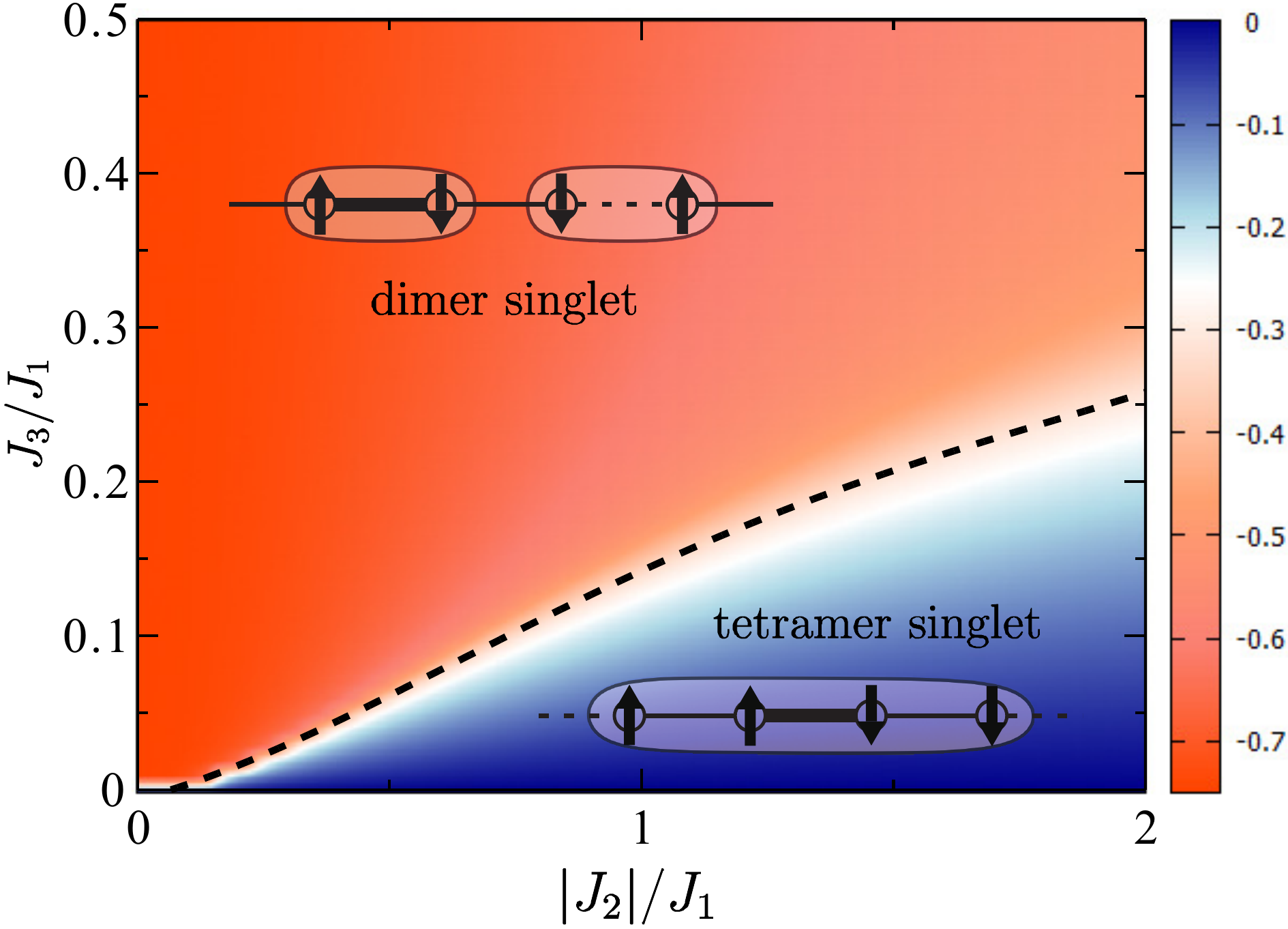}
  \caption{Ground-state Phase diagram with $|J_2|/J_1$ and $J_3/J_1$. The dashed line is the phase boundary between the tetramer-singlet and dimer-singlet states, estimated from the central charge. The color map displays the spin-spin correlation between two neighboring spins on $J_3$ bond. Schematic picture for the each state is also shown.
  }
  \label{phasediagram}
\end{figure}

\subsection{Phase diagram}

Fig.~\ref{phasediagram} shows the ground-state phase diagram of system \eqref{ham}. The phase boundary between the tetramer-singlet and dimer-singlet phases has been determined based on the results of central charge. At $|J_2|=J_3=0$, only singlet $J_1$-bonds as well as uncoupled spins exist. The tetramer-singlet state is stabilized if FM $J_2$ is switched on; and the dimer-singlet state is stabilized if AFM $J_3$ is switched on. In the small $|J_2|/J_1$ regime, therefore, a phase transition between the two singlet states is driven by the competition between $J_2$ and $J_3$. By comparing the energies of isolated tetramer singlet and isolated dimer singlet, we obtain a relation $J_3/J_1=(|J_2|/J_1)^2/2$ giving the phase boundary. Thus, the boundary line, i.e., critical $J_3/J_1$ value rises slowly with increasing $|J_2|/J_1$. More details are discussed in in Appendix A.

Whereas in the large $|J_2|/J_1$ regime, the system \eqref{ham} can be effectively mapped onto a spin-$1$ Heisenberg chain with alternating bonds $J_1/4$ and $J_3/4$. This mapping becomes exact in the limit of $|J_2|=\infty$. In this limit, the system is in either the tetramer-singlet or dimer-singlet state depending on the ratio between $J_1$ and $J_3$. The tetramer-singlet and dimer-singlet states correspond to the (2,0)- and (1,1)-type valence-bond-solid (VBS) states~\cite{Totsuka95}, respectively. Since the critical ratio between the (2,0)- and (1,1)-type VBS states has been estimated to be $J_3/J_1=0.58736$ in the spin-$1$ chain~\cite{Nakamura02,Miyakoshi16}, the phase boundary is expected to saturate at $J_3/J_1=0.58736$ in the large $|J_2|/J_1$ regime of system \eqref{ham}. In fact, a critical value $J_3/J_1=0.58632$ is obtained at $|J_2|/J_1=1000$. More details are discussed in Appendix A.

Fig.~\ref{phasediagram} also presents the spin-spin correlation between two neighboring sites on $J_3$ bond. In the tetramer-singlet phase it is nearly zero since four spins in each tetramer are almost screened. In the dimer-singlet phase it would be close to $-3/4$. We find that the spin-spin correlation rapidly changes around the phase boundary. It is interesting that the phase boundary is roughly coincident with a line of the spin-spin correlation $-1/4$ meaning a strong competition between the two singlet states. A similar trend was pointed out by cluster mean field calculations in Ref.~\onlinecite{Singhania18}.

\begin{figure}
  \includegraphics[width=0.8\columnwidth]{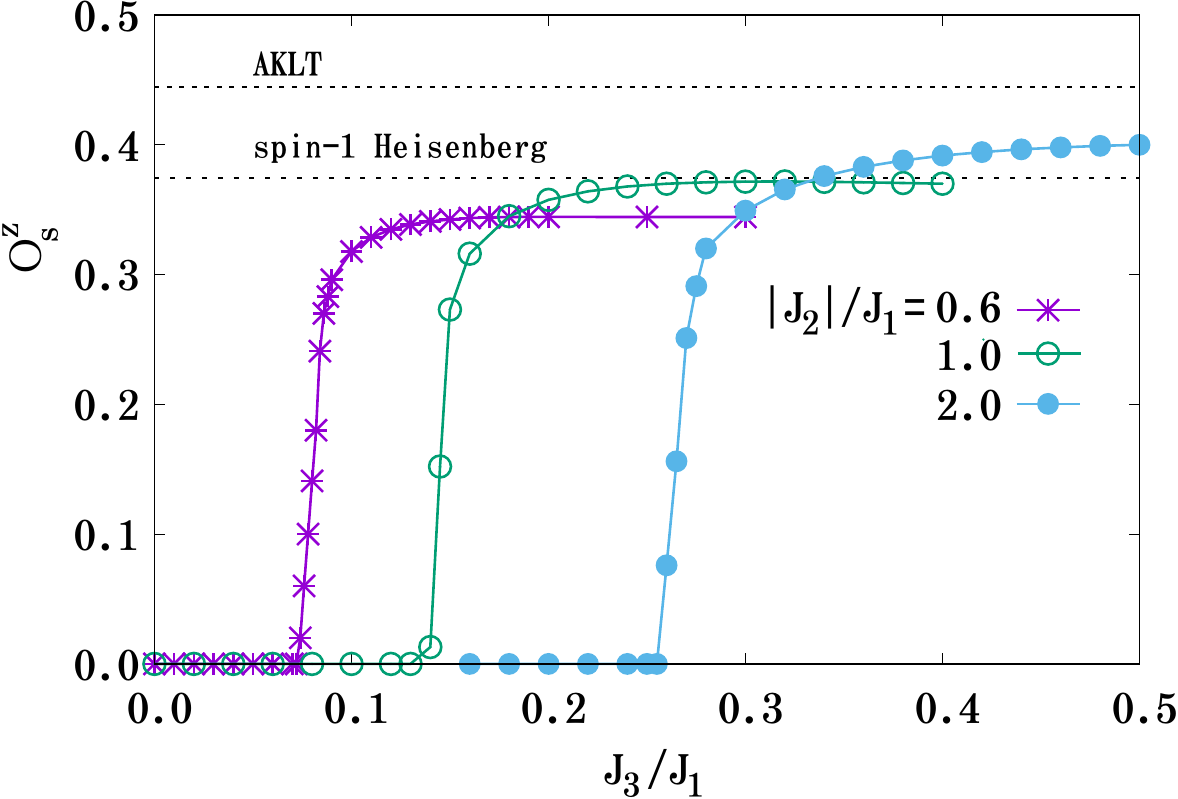}
  \caption{String order parameter as a function of $J_3/J_1$ for $|J_2|/J_1=0.6$, $1$ and $2$. The dotted lines indicate the values for the AKLT model and the spin-$1$ Heisenberg chain.
  }
  \label{stringorder}
\end{figure}

\subsection{String order}

If two spin-$\frac{1}{2}$'s on the $J_2$ bond are contracted to an effective spin-$1$ degree of freedom, finite spin gap in the dimer-singlet state might be interpreted as a Haldane gap. In a Haldane system, a topological order characterized by the hidden $\mathbb{Z}_2 \times \mathbb{Z}_2$ symmetry breaking is naively expected. To investigate the possibility of the hidden order, we examine the string order parameter:
\begin{align}
\nonumber
\mathcal O_\mathrm{s}^z = - \lim_{|i-j| \to \infty} &\langle (S^z_{i,3}+S^z_{i,4})\\
&\times\exp(i\pi\sum_{k=i+1}^{j-1} \sum_{l=1}^4 S^z_{k,l})(S^z_{j,1}+S^z_{j,2}) \rangle.
\label{stringorder1}
\end{align}
For our system \eqref{ham}, Eq. \eqref{stringorder1}) can be simplified as
\begin{align}
\nonumber
\mathcal O_\mathrm{s}^z = - \lim_{|i-j| \to \infty} (-4)^{2(j-i-1)} &\langle (S^z_{i,3}+S^z_{i,4})\\
&\prod_{k=i+1}^{j-1}\prod_{l=1}^4 S^z_{k,l} (S^z_{j,1}+S^z_{j,2}) \rangle.
\label{stringorder2}
\end{align}
In Eqs.~\eqref{stringorder1} and \eqref{stringorder2}, although the sites $3$-$4$ in $i$-th tetramer and sites $1$-$2$ in $j$-th tetramer are chosen as the $J_2$-bonded spin-$\frac{1}{2}$ pairs, i.e., the effective spin-$1$ sites, the results does not depend on the choice; sites $1$-$2$ in $i$-th tetramer and sites $1$-$2$ in $j$-th tetramer, sites $3$-$4$ in $i$-th tetramer and sites $3$-$4$ in $j$-th tetramer.

In Fig.~\ref{stringorder} the string order parameter in the thermodynamic limit is plotted as a function of $J_3/J_1$ for several $|J_2|/J_1$ values. The finite value of $\mathcal O_\mathrm{s}^z$ suggests the formation of a singlet state with a hidden topological long-range order. Interestingly, the string order starts to develop rapidly at the QCP from tetramer-singlet to dimer-singlet phases but it increases continuously from $0$ as a consequence of the second-order transition between the two singlet states. It seems to saturate quickly with $J_3/J_1$. The saturation value is increased with increasing $|J_2|/J_1$ because the Haldane's VBS picture becomes more complete for larger $|J_2|/J_1$.
Already at $|J_2|/J_1=1$ the value is close to $\mathcal O_\mathrm{s}^z\simeq0.3743$ for the spin-$1$ Heisenberg chain. In the limit $|J_2|/J_1\to\infty$ it approaches $O_\mathrm{s}^z=\frac{4}{9}\simeq0.4444$ for the {\it perfect} VBS state for the Affleck-Kennedy-Lieb-Tasaki (AKLT) model~\cite{Affleck87}. Note that with further increasing $J_3/J_1$ the string order goes down to $0$ at some $J_3/J_1>1$ because the system goes again into the tetramer-singlet state where a $J_3$ bond with the neighboring $J_2$ bonds forms the tetramer singlet state.

\subsection{Magnetization with external field}

\begin{figure}
  \includegraphics[width=0.85\columnwidth]{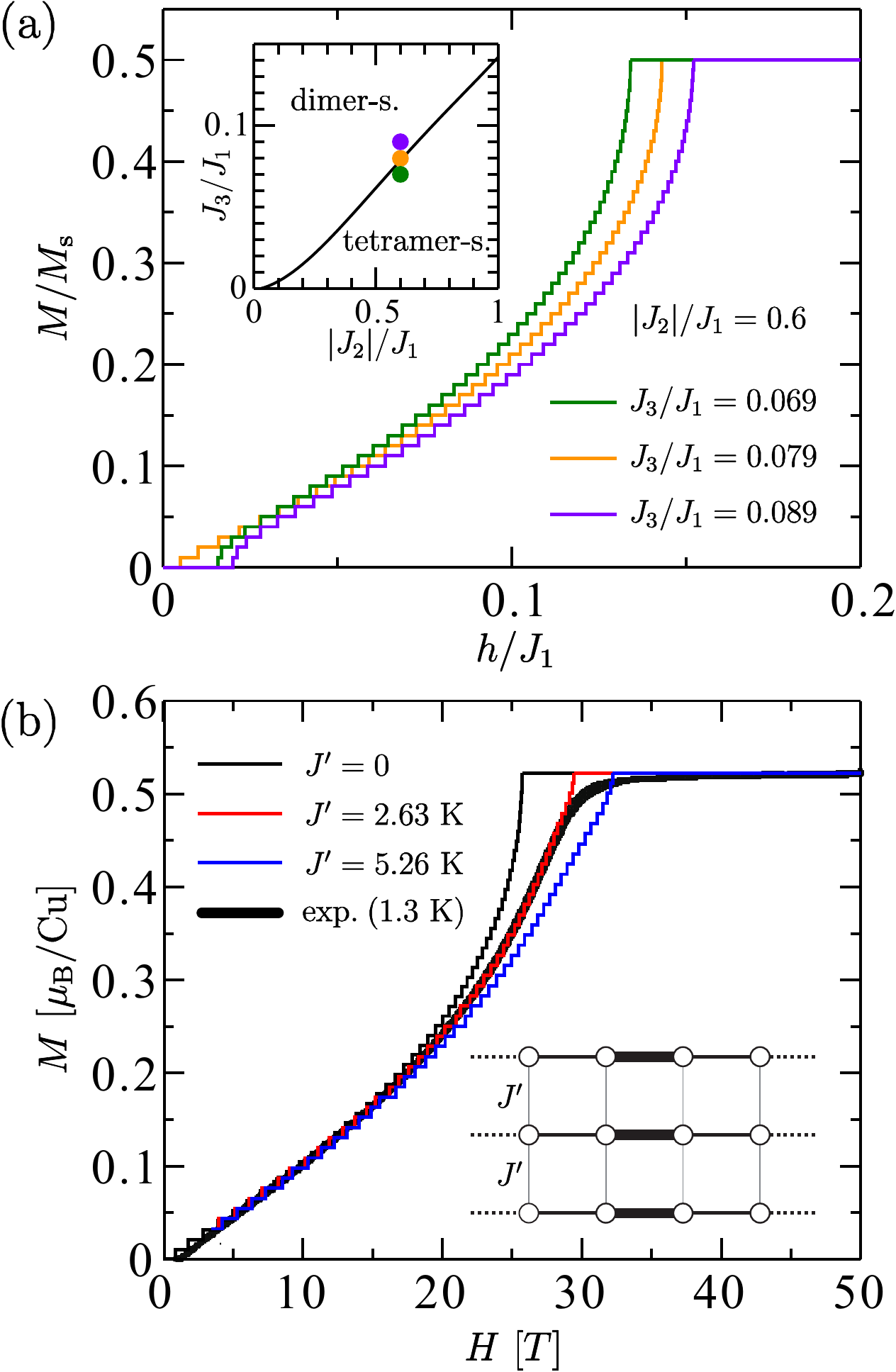}
  \caption{(a) Magnetization $M/M_{\rm s}$ of the system \eqref{ham} near the phase transition calculated by DMRG, where $M_{\rm s}$ is the full-saturation value. The inset indicates the location of used parameters in the phase diagram. (b) Fitting of the experimental magnetization curve~\cite{Hase16} with a three-leg tetramer chains coupled by perpendicular interchain interaction $J^\prime$. The inset shows the lattice structure of three-leg tetramer chains.
  }
  \label{magnetization}
\end{figure}

The experimental magnetization of CuInVO$_5$ at $K=1.3$ K begins to increase almost linearly from zero field~\cite{Hase16}, as seen in Fig.~\ref{magnetization}(b). This indicates that the system is in a gapless or tiny-gapped state at $T=0$. Thus, we consider the magnetization $M$ with external field $h$ around the phase boundary. According to Ref.~\cite{Hase16}, we here focus on $|J_2|/J_1=0.6$ where the critical $J_3/J_1$ is $0.079$. In Fig.~\ref{magnetization}(a) the magnetization curves of system \eqref{ham} near the critical point are plotted. At the critical point $J_3/J_1=0.079$, the magnetization is smoothly connected to $M=0$ with approaching $h=0$, being consistent with the experimental observation. However, when $J_3/J_1$ deviates only by $\pm0.01$ from the critical value, the system enters into the singlet phases and consequently the gapped features are clearly visible in the magnetization process. Its excitation gap corresponds to $H\sim4$ T in CuInVO$_5$. Therefore, we argue that CuInVO$_5$ stands very close to the QCP. Furthermore, with $|J_2|/J_1=0.6$ fixed we see a significant discrepancy between our critical $J_3/J_1=0.079$ and the mean-field value ($J_3/J_1=0.125$) leading to a `gapless' magnetization in Ref.~\onlinecite{Hase16}. It means that the quantum fluctuations, which are strongly suppressed in the mean-field calculation, play a crucial role in the phase transition between two singlet states.

The magnetization of system \eqref{ham} also exhibits a divergent increase near the half-saturation $M/M_{\rm s}=0.5$. This is a typical feature of 1D Heisenberg systems. A similar behavior is found at any point on the phase boundary (see Appendix C). But the experimental magnetization near the half-saturation is more gentle which in general, can be effected by AFM interchain coupling. Although the structure of interchain couplings for CuInVO$_5$ is unknown, the dominant one should be unfrustrated and AFM because a N\'eel order has been experimentally observed. We then simply assume a perpendicular AFM interchain coupling $J^\prime$. To examine the effect of $J^\prime$, we employ three-leg tetramer chains [see the inset of Fig.~\ref{magnetization}(b)] to maintain the gapless feature. A fitting of the experimental magnetization curve using the three-leg tetramer chains is shown in Fig.~\ref{magnetization}(b). We can see a good agreement by assuming only $1\%$ AFM interchain coupling of $J_1$: The used parameters are $J_1=263$ K, $J_2=-158$ K, $J_3=21$ K, and $J^\prime=2.63$ K. We note that this fitting is not unique because a comparable agreement can be also achieved even with the other $|J_2|/J_1$ values. Nevertheless, in either case the interchain coupling is likely to be very weak.

\begin{figure}
  \includegraphics[width=0.8\columnwidth]{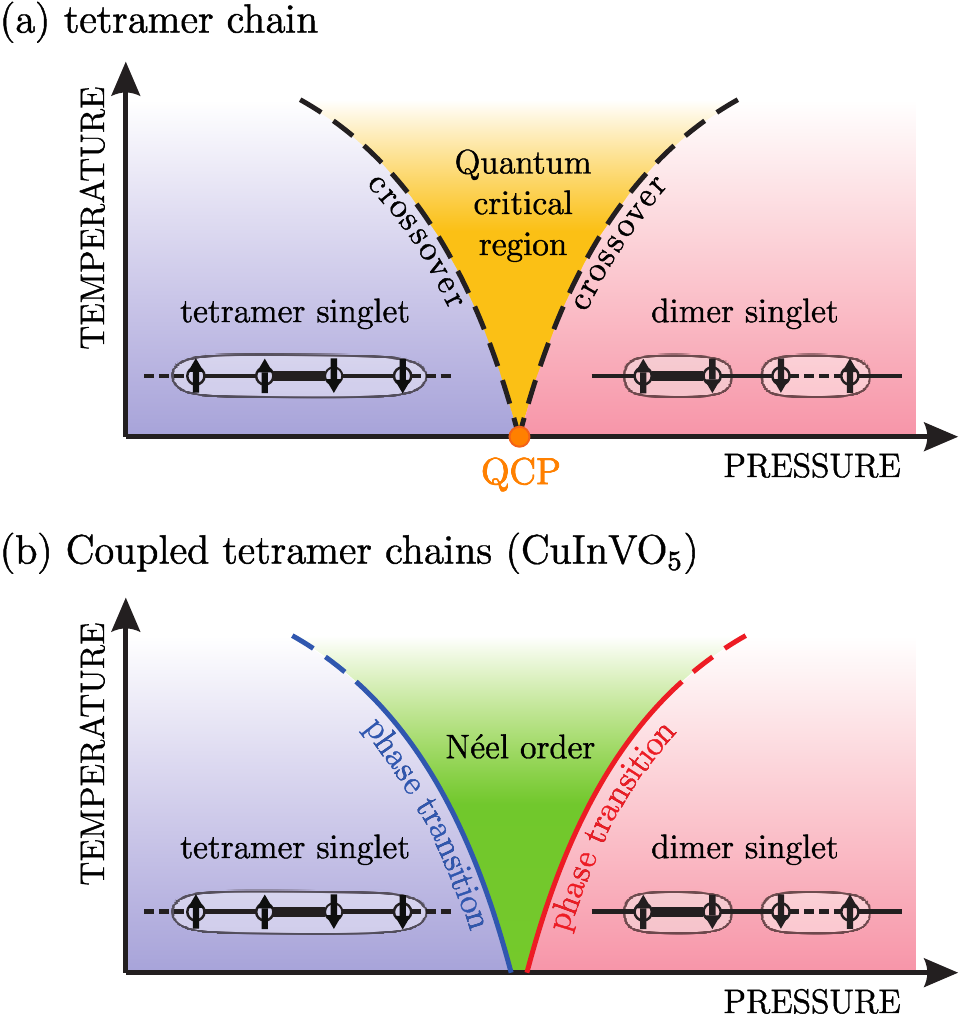}
  \caption{Schematic phase diagram of CuInVO$_5$ based on the theoretical speculation with (a) tetramer chain and (b) coupled tetramer chains. At zero temperature, when the critical tetramer chains are connected by AFM interchain coupling, only a narrow region near the QCP is substituted by a N\'eel phase; then, the N\'eel phase extends to nonzero temperature. The dotted and solid lines between different regions denote crossover and phase transition, respectively.
}
  \label{speculation}
\end{figure}

\section{Conclusion and Discussion}
We studied the 1D spin-$\frac{1}{2}$ Heisenberg model consisting of coupled tetramers using the DMRG method. Based on the results of spin gap and central charge, we mapped out the ground-state phase diagram as a function of intra-tetramer coupling and inter-tetramer coupling. Depending on coupling ratio, we found two singlet phases in the phase diagram: tetramer-singlet state where each tetramer forms a singlet unit; dimer-singlet state where each dimer forms a singlet pair. Interestingly, the dimer-singlet state is interpreted as a Haldane state with the hidden $\mathbb{Z}_2 \times \mathbb{Z}_2$ symmetry breaking. We also showed that the spin gaps in the both singlet phases continuously approach to zero and the central charge is unity  at the phase boundary. This defines a QCP where a second-order transition occurs between the two singlet phases. By analyzing the experimental magnetization curve, we argued that the possible exchange coupling parameters of CuInVO$_5$ are close to the QCP at ambient pressure; and the interchain coupling needed to realize the N\'eel order is very small, only $\sim1\%$ of the AFM intra-tetramer coupling.

Lastly, we provide some speculations about the experimental realization in CuInVO$_5$. As demonstrated above, the tetramer chain \eqref{ham} undergoes a second-order transition between tetramer-singlet and dimer-singlet phases at $T=0$. The 1D nature of CuInVO$_5$ would be well described by the single tetramer chain near the QCP; however, the observed N\'eel order is never achieved as long as the single chain is considered [see Fig.~\ref{speculation}(a)]. And yet, when the (nearly) critical tetramer chains are connected by weak AFM interchain coupling, the phase diagram is changed as illustrated in Fig.~\ref{speculation}(b); namely, the QCP and quantum critical region at low temperature are replaced by N\'eel phase, just as a N\'eel order can be realized in the coupled critical Heisenberg chains~\cite{Yasuda05}; on the other hand, the two singlet phases mostly remain as they are since a singlet state is robust against additional couplings. The N\'eel temperature might scale to the magnitude of AFM interchain coupling. In Ref.~\onlinecite{Hase16}, the N\'eel order might be observed somewhere in the green range of Fig.~\ref{speculation}(b). If such is the case, by applying pressure or stretch, varying the balance of exchange interactions, a transition from N\'eel to singlet phases could be driven. More intriguing expectation may be that a two-step transition from paramagnetic to singlet via N\'eel states could be observed by lowering temperature. However, it depends on how the N\'eel phase extends at finite temperature. It is to be hoped that future theoretical and experimental researches will clarify this point.

\section*{Acknowledgements}

We thank U. Nitzsche for technical assistance. This work is supported by ARC under the grant number DP180101483 and SFB 1143 of the Deutsche Forschungsgemeinschaft. Computations were carried out on the ITF/IFW Dreden, Germany.

\appendix

\section{Finite-size scaling}

\begin{figure}[b]
  \includegraphics[width=1.0\columnwidth]{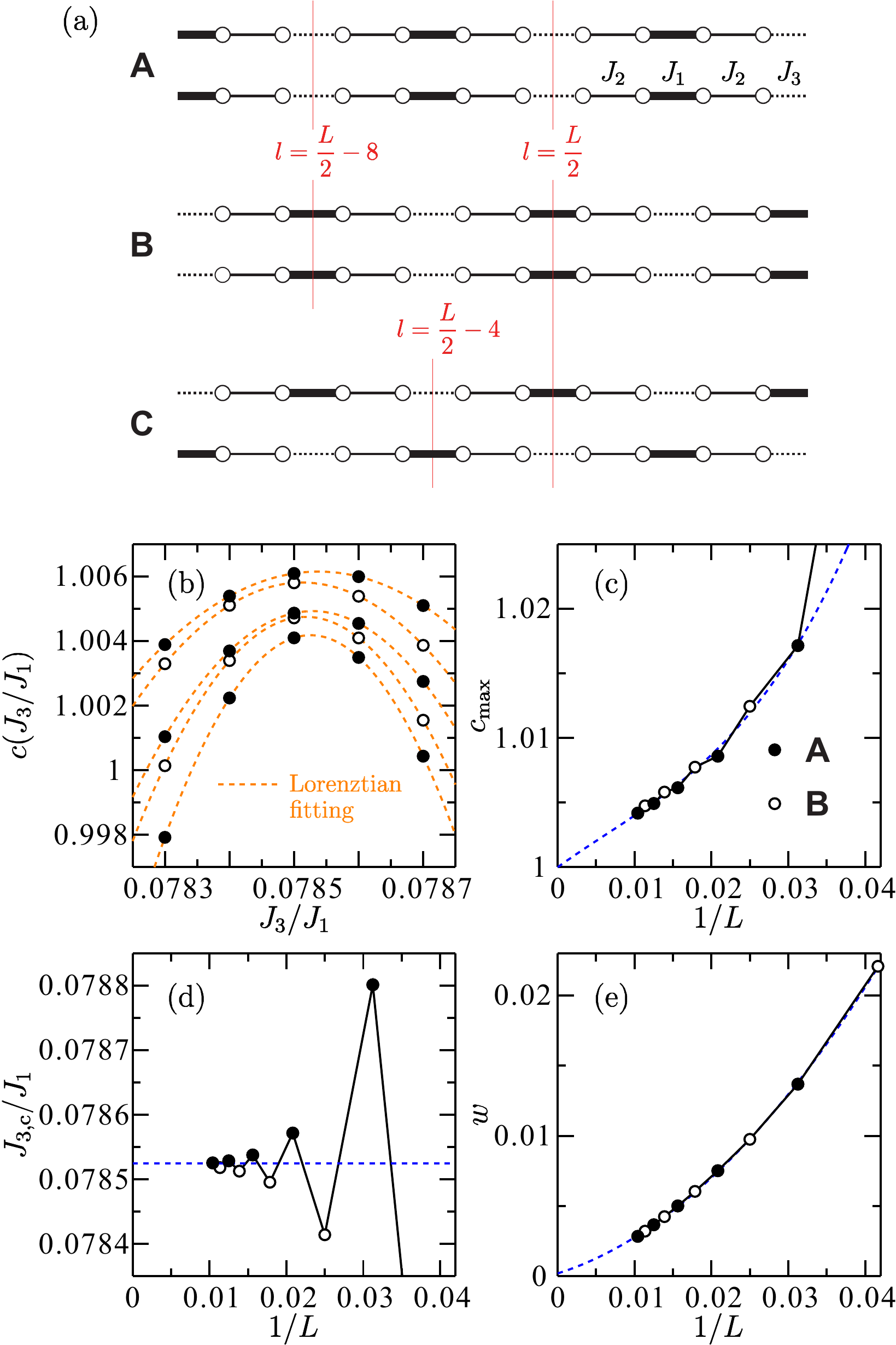}
  \caption{(a) Three kinds of partitioning of the periodic system in our calculation of central charge. (b) Fitting of the central charge with fixed system size by the Lorentzian function~\eqref{lolentzian}. The system size is $L=96$, $88$, $80$, $72$, and $64$ from bottom to top. Finite-size scaling analyses of (c) the peak height $c_{\rm max}$, (d) position $J_{3,{\rm c}}/J_1$, and (e) half-width $w$ as a function of $1/L$. The blue dotted line denotes a scaling function.
  }
  \label{c_scaling}
\end{figure}

In Fig.~2(b) of the main text the central charge $c$ is plotted as a function of $J_3/J_1$ at fixed $|J_2|/J_1$ values for several system lengths. A single Lorentzian-like peak is obtained in each case. Apparently, the peak height is $c\approx1$, the peak position is almost independent of the system size, and the peak width is decreased with increasing the system size; therefore, this peak seems to indicate a quantum critical point (QCP) between the two singlet phases. However, if the peak indeed corresponds to a single QCP, the following conditions must be fulfilled: (i) The peak height goes to $1$, (ii) the peak width is shrunk to $0$, i.e., $\delta$-peak, and (iii) the peak position converges to a finite value in the thermodynamic limit $L\to\infty$.

To confirm it, the finite-size scaling analysis for $|J_2|/J_1=0.6$ is performed in Fig.~\ref{c_scaling}. First, as shown in Fig.~\ref{c_scaling}(b), for each the system size, the peak height ($c_{\rm max}$), position ($J_{3,{\rm c}}/J_1$), and half-width ($w$) are estimated by fitting several points in a narrow $J_3/J_1$ region around the maximum with a Lorentzian function
\begin{align}
c(J_3/J_1)=\frac{c_{\rm max}w^2}{(J_3-J_{3,{\rm c}})^2/J_1^2+w^2}.
\label{lolentzian}
\end{align}
The values of $c_{\rm max}$, $J_{3,{\rm c}}/J_1$, and $w$ are determined as fitting parameters for each the peak. Using the determined values, finite-size scaling analyses of the peak height, position, and half-width as a function of $1/L$ are shown in Fig.~\ref{c_scaling}(c-e), respectively We clearly see that the values are extrapolated to $c_{\rm max}=1$, $J_{3,{\rm c}}/J_1=0.078525$, and $w=0$ in the thermodynamic limit $L\to\infty$. We thus confirm that the QCP at $J_3/J_1=0.078525$ is indicated by a $\delta$-peak of the central charge with height $1$.

Let now us comment on the oscillation of $c_{\rm max}$, $J_{3,{\rm c}}/J_1$, and $w$ with the system size, seen in Fig.~\ref{c_scaling}(c-e). The oscillation is caused by a technical reason: The system is divided into two subsystems when we calculate the entanglement entropy $S_L(\frac{L}{2})$ and $S_L(\frac{L}{2}-8)$ for obtaining the central charge via Eq.~(3) of the main text. The cut position (bond) depends on how to construct the DMRG block. In our calculations there are two possibilities of appropriate cut; one is to cut two $J_3$ bonds and the other is to cut two $J_1$ bonds as shown in {\bf A} and {\bf B} of Fig.~\ref{c_scaling}(a), respectively. In Fig.~\ref{c_scaling}(b-e), the values obtained with cut {\bf A} and {\bf B} are plotted as filled and open circles, respectively. This oscillation is not very crucial for the scaling analysis in this study. But truthfully, the scaling analysis should be performed separately between the cases of {\bf A} and {\bf B}. For some parameters, we also explored another cut manner as {\bf C} in Fig.~\ref{c_scaling}(a); where the system is divided at each of $J_1$ and $J_3$ bonds. In fact, this cut manner leads to a faster convergence to the thermodynamic limit. It is so because that the central charge can be obtained from the 4-th neighbor entanglement entropys $S_L(\frac{L}{2})$ and $S_L(\frac{L}{2}-4)$ in the {\bf C} scheme instead of the 8-th neighbors $S_L(\frac{L}{2})$ and $S_L(\frac{L}{2}-8)$ in the {\bf A} and {\bf B} schemes.

\section{phase boundary for limiting cases}

Let us start with an isolated tetramer limit  ($J_2<0$, $J_3=0$). The ground-state energy of isolated tetramer is
\begin{align}
\frac{E_{\rm tet}}{J_1}=-\frac{1}{4}\left(1+\frac{2J_2}{J_1}+2\sqrt{1-\frac{2J_2}{J_1}+\left(\frac{2J_2}{J_1}\right)^2}\right).
\label{tetramerenergy}
\end{align}
In the limit $|J_2| \ll J_1$, Eq.\eqref{tetramerenergy} is approximated by
\begin{align}
\frac{E_{\rm tet}}{J_1}=-\frac{3}{4}\left[1+\left(\frac{J_2}{J_1}\right)^2\right].
\label{tetramerenergy2}
\end{align}
Whereas in isolated dimer limit ($J_2=0$, $J_3>0$), the ground-state energy per tetramer is a simple sum of two singlet pairs on the $J_1$ and $J_3$ bonds:
\begin{align}
\frac{E_{\rm dim}}{J_1}=-\frac{3}{4}\left(1+\frac{J_3}{J_1}\right).
\label{dimersenergy}
\end{align}

\begin{figure}[t]
  \includegraphics[width=1.0\columnwidth]{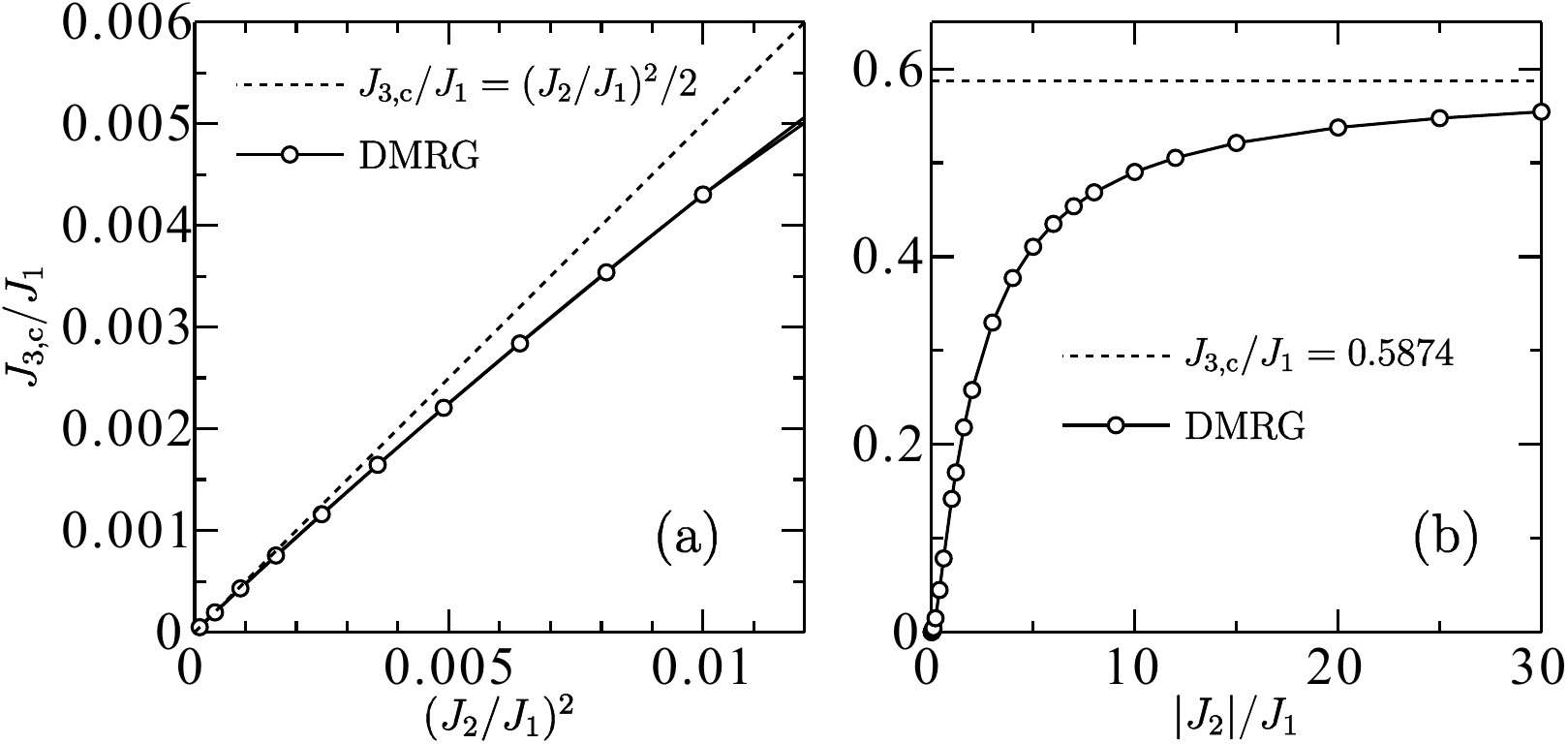}
  \caption{Critical $J_3/J_1$ value of the tetramer chain as a function of $|J_2|/J_1$ for (a) small $|J_2|/J_1$ region and (b) wide range of $|J_2|/J_1$. The analytical asymptotic lines are also plotted as dotted lines.
  }
  \label{critical_limit}
\end{figure}

By comparing Eqs.~\eqref{tetramerenergy2} and \eqref{dimersenergy} with taking into consideration that the number of $J_2$ bonds are twice as many as that of $J_3$ bonds in the system, we obtain
\begin{align}
\frac{J_3}{J_1}=\frac{1}{2}\left(\frac{J_2}{J_1}\right)^2.
\label{smallJ2J3boundary}
\end{align}
The phase boundary is given by Eq.~\eqref{smallJ2J3boundary} at the small $|J_2|$ and $J_3$ region. In Fig.~\ref{critical_limit}(a) we plot the critical values of $J_3/J_1$ obtained from the central charge as a function of $(J_2/J_1)^2$. We can see that the numerical values approach asymptotically to the analytical line \eqref{smallJ2J3boundary} with decreasing $|J_2|$ and $J_3$.

In the large $|J_2|$ limit, two spin-$\frac{1}{2}$'s on each $J_2$ bond form a spin-triplet pair. By relating three states $|\uparrow\uparrow\rangle$, $|\uparrow\downarrow\rangle+|\downarrow\uparrow\rangle)/\sqrt{2}$, and $|\downarrow\downarrow\rangle$ to $S^z=1$, $0$, and $-1$ states, respectively, the resultant spin on the $J_2$ bond can be reduced to a spin-$1$ degree of freedom. Therefore, the tetramer chain can be effectively mapped onto a spin-$1$ Heisenberg chain with bond alternation:
\begin{align}
\mathcal{H}=\sum_{i=1}^{L/4}\left(\frac{J_1}{4}\vec{S}^{(1)}_{2i-1}\cdot\vec{S}^{(1)}_{2i}+\frac{J_3}{4}\vec{S}^{(1)}_{2i}\cdot\vec{S}^{(1)}_{2i+1}\right)+{\rm const.},
\label{S1ham}
\end{align}
where $\vec{S}^{(1)}_i$ is the spin-$1$ operator at site $i$. Then, the tetramer-singlet and dimer-singlet states correspond to the (2,0)- and (1,1)-type valence-bond-solid (VBS) states, respectively. The critical value between two VBS states has been estimated as $J_3/J_1=0.58736$. As shown in Fig.~\ref{critical_limit}(b), the critical $J_3/J_1$ value of the tetramer chain approaches asymptotically to $0.58736$ in the large $|J_2|$ regime.

\begin{figure}
  \includegraphics[width=1.0\columnwidth]{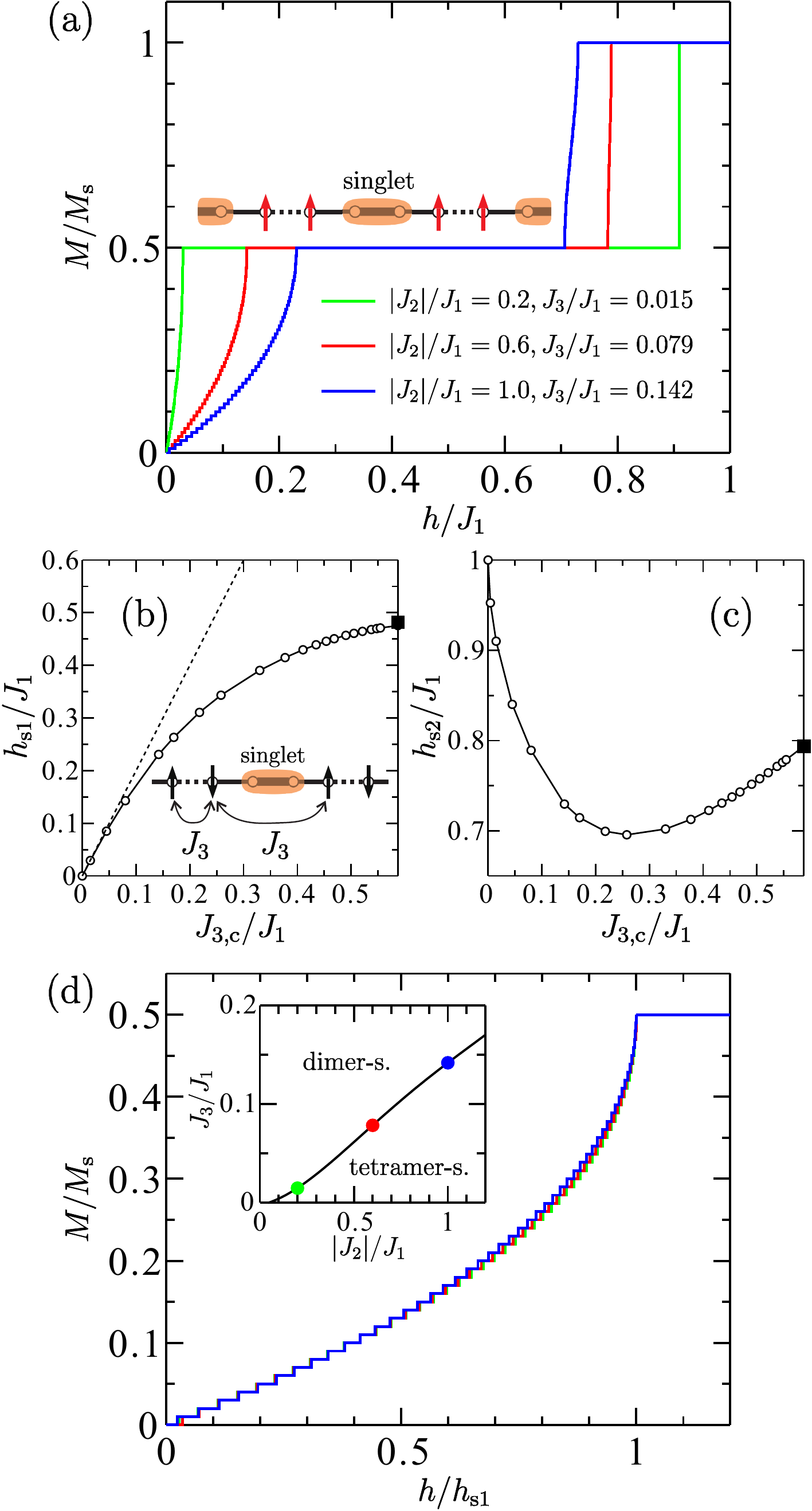}
  \caption{(a) Magnetization as a function of external field for several parameter sets on the phase boundary between two singlet states. A schematic spin structure on the 1/2-plateau is shown in the inset. Magnitude of the external field leading to the (b) half-saturation and (c) full-saturation as as function of the critical value of $J_3/J_1$. (d) Magnetization as a function of normalized external field by $h_{\rm s1}$ for several parameters on the phase boundary. The inset shows the positions of the parameter sets in the phase diagram.
  }
  \label{magnetization_supl}
\end{figure}

\section{Magnetization}

We here explain the general properties of magnetization just on the phase boundary between two singlet states. For relevance to the experimental observation, we plot only the low-field part of magnetization curve in the main text. In Fig.~\ref{magnetization_supl}(a) we show a whole picture of the magnetization curve, including the full-saturation $M/M_{\rm s}=1$ for some parameter sets on the phase boundary . In each case the gapless behavior is clearly seen near zero field. With increasing the external field, we find a wide plateau at $M/M_{\rm s}=1/2$, where each the $J_1$ bond form a singlet pair and the other spins are polarized along the field direction. 

We then consider the saturation fields. Hereafter, the magnitudes of external field where the magnetization reaches the half-saturation $M/M_{\rm s}=1/2$ and full-saturation $M/M_{\rm s}=1$ are denoted as $h_{\rm s1}$ and $h_{\rm s2}$, respectively. In Fig.~\ref{magnetization_supl}(b) the half-saturation field $h_{\rm s1}$ is plotted as a function of critical $J_3$ value ($\equiv J_{\rm 3,c}$). When $|J_2|$ and $J_3$ are small, at the QCP it would be fair to assume that $J_3$ is counterbalanced to the interaction between two spins on the sides of $J_1$ bond, as sketched in the inset of Fig.~\ref{magnetization_supl}(b). Therefore, the system might be mapped onto a spin-$\frac{1}{2}$ Heisenberg chain with uniform interaction $J_{\rm 3,c}$ by renormalizing the spin degrees of freedom on the $J_1$ bond into the {\it virtual} interaction $J_3$. The saturation field of this chain is $h_{\rm s1}=2J_{\rm 3,c}$; as seen in Fig.~\ref{magnetization_supl}(b), this relation gives a good approximation for the half-saturation field at small  $|J_2|$ and $J_3$ region. The full-saturation field $h_{\rm s2}/J_1$ is plotted as a function of $J_{3,c}$ in Fig.~\ref{magnetization_supl}(c). It is $1$ in the small $|J_2|$ and $J_3$ limit is $1$ because the system consists only of singlet dimers on the $J_1$ bond. When $|J_2|$ and $J_3$ are introduced, singlet-triplet splitting of the singlet dimers is narrowed so that $h_{\rm s,2}/J_1$ is decreased. Whereas in the large $|J_2$ limit, the system can be mapped onto a spin-1 Heisenberg chain with alternating bonds $J_1/4$ and $J_3/4$. The half- and full-saturation fields of this spin-$1$ chain at the critical point $J_3/J_1=0.58736$ between the (2,0)- and (1,1)-type VBS states are estimated as $h_{\rm s,1}=0.48191297$ and $h_{\rm s,2}=0.7936825$, respectively. Hence, the values of $h_{\rm s,1}/J_1$ and  $h_{\rm s,2}/J_1$ of the tetramer chain approach $0.48191297$ and $0.7936825$, respectively, at $J_{\rm 3,c}/J_1=0.58736$ ($|J_2|\to\infty$). 

Finally, let us check the parameter dependence of magnetization curve at the QCP. Fig.~\ref{magnetization_supl}(d) shows the magnetization curve for the tetramer chain as a function of normalized external field by $h_{\rm s,1}$ for several critical points. The shape of magnetization curve is almost independent of used parameters in the realistic range.  Thus, a unique fitting of the experimental magnetization curve within the tetramer chain would be difficult.

\bibliography{CuInVO5}

\begin{thebibliography}{44}%
\makeatletter
\providecommand \@ifxundefined [1]{%
 \@ifx{#1\undefined}
}%
\providecommand \@ifnum [1]{%
 \ifnum #1\expandafter \@firstoftwo
 \else \expandafter \@secondoftwo
 \fi
}%
\providecommand \@ifx [1]{%
 \ifx #1\expandafter \@firstoftwo
 \else \expandafter \@secondoftwo
 \fi
}%
\providecommand \natexlab [1]{#1}%
\providecommand \enquote  [1]{``#1''}%
\providecommand \bibnamefont  [1]{#1}%
\providecommand \bibfnamefont [1]{#1}%
\providecommand \citenamefont [1]{#1}%
\providecommand \href@noop [0]{\@secondoftwo}%
\providecommand \href [0]{\begingroup \@sanitize@url \@href}%
\providecommand \@href[1]{\@@startlink{#1}\@@href}%
\providecommand \@@href[1]{\endgroup#1\@@endlink}%
\providecommand \@sanitize@url [0]{\catcode `\\12\catcode `\$12\catcode
  `\&12\catcode `\#12\catcode `\^12\catcode `\_12\catcode `\%12\relax}%
\providecommand \@@startlink[1]{}%
\providecommand \@@endlink[0]{}%
\providecommand \url  [0]{\begingroup\@sanitize@url \@url }%
\providecommand \@url [1]{\endgroup\@href {#1}{\urlprefix }}%
\providecommand \urlprefix  [0]{URL }%
\providecommand \Eprint [0]{\href }%
\providecommand \doibase [0]{https://doi.org/}%
\providecommand \selectlanguage [0]{\@gobble}%
\providecommand \bibinfo  [0]{\@secondoftwo}%
\providecommand \bibfield  [0]{\@secondoftwo}%
\providecommand \translation [1]{[#1]}%
\providecommand \BibitemOpen [0]{}%
\providecommand \bibitemStop [0]{}%
\providecommand \bibitemNoStop [0]{.\EOS\space}%
\providecommand \EOS [0]{\spacefactor3000\relax}%
\providecommand \BibitemShut  [1]{\csname bibitem#1\endcsname}%
\let\auto@bib@innerbib\@empty
\bibitem [{\citenamefont {Coleman}\ and\ \citenamefont
  {Schofield}(2005)}]{Coleman05}%
  \BibitemOpen
  \bibfield  {author} {\bibinfo {author} {\bibfnamefont {P.}~\bibnamefont
  {Coleman}}\ and\ \bibinfo {author} {\bibfnamefont {A.~J.}\ \bibnamefont
  {Schofield}},\ }\bibfield  {title} {\bibinfo {title} {Quantum criticality},\
  }\href@noop {} {\bibfield  {journal} {\bibinfo  {journal} {Nature}\ }\textbf
  {\bibinfo {volume} {433}},\ \bibinfo {pages} {226} (\bibinfo {year}
  {2005})}\BibitemShut {NoStop}%
\bibitem [{\citenamefont {Sachdev}\ and\ \citenamefont
  {Keimer}(2011)}]{Sachdev11}%
  \BibitemOpen
  \bibfield  {author} {\bibinfo {author} {\bibfnamefont {S.}~\bibnamefont
  {Sachdev}}\ and\ \bibinfo {author} {\bibfnamefont {B.}~\bibnamefont
  {Keimer}},\ }\bibfield  {title} {\bibinfo {title} {Quantum criticality},\
  }\href {https://doi.org/10.1063/1.3554314} {\bibfield  {journal} {\bibinfo
  {journal} {Physics Today}\ }\textbf {\bibinfo {volume} {64}},\ \bibinfo
  {pages} {29} (\bibinfo {year} {2011})}\BibitemShut {NoStop}%
\bibitem [{\citenamefont {Moriya}(1985)}]{Moriya85}%
  \BibitemOpen
  \bibfield  {author} {\bibinfo {author} {\bibfnamefont {T.}~\bibnamefont
  {Moriya}},\ }\href {https://books.google.de/books?id=t8HvAAAAMAAJ} {\emph
  {\bibinfo {title} {Spin fluctuations in itinerant electron magnetism}}},\
  Springer series in solid-state sciences\ (\bibinfo  {publisher}
  {Springer-Verlag},\ \bibinfo {year} {1985})\BibitemShut {NoStop}%
\bibitem [{\citenamefont {Binney}\ \emph {et~al.}(1992)\citenamefont {Binney},
  \citenamefont {Binney}, \citenamefont {Binney}, \citenamefont {Dowrick},
  \citenamefont {Fisher}, \citenamefont {Newman},\ and\ \citenamefont
  {Newman}}]{Binney92}%
  \BibitemOpen
  \bibfield  {author} {\bibinfo {author} {\bibfnamefont {J.}~\bibnamefont
  {Binney}}, \bibinfo {author} {\bibfnamefont {M.}~\bibnamefont {Binney}},
  \bibinfo {author} {\bibfnamefont {J.}~\bibnamefont {Binney}}, \bibinfo
  {author} {\bibfnamefont {N.}~\bibnamefont {Dowrick}}, \bibinfo {author}
  {\bibfnamefont {A.}~\bibnamefont {Fisher}}, \bibinfo {author} {\bibfnamefont
  {M.}~\bibnamefont {Newman}},\ and\ \bibinfo {author} {\bibfnamefont
  {A.}~\bibnamefont {Newman}},\ }\href
  {https://books.google.de/books?id=lvZcGJI3X9cC} {\emph {\bibinfo {title} {The
  Theory of Critical Phenomena: An Introduction to the Renormalization
  Group}}},\ Oxford Science Publ\ (\bibinfo  {publisher} {Clarendon Press},\
  \bibinfo {year} {1992})\BibitemShut {NoStop}%
\bibitem [{\citenamefont {Timusk}\ and\ \citenamefont
  {Statt}(1999)}]{Timusk99}%
  \BibitemOpen
  \bibfield  {author} {\bibinfo {author} {\bibfnamefont {T.}~\bibnamefont
  {Timusk}}\ and\ \bibinfo {author} {\bibfnamefont {B.}~\bibnamefont {Statt}},\
  }\bibfield  {title} {\bibinfo {title} {The pseudogap in high-temperature
  superconductors: an experimental survey},\ }\href
  {https://doi.org/10.1088/0034-4885/62/1/002} {\bibfield  {journal} {\bibinfo
  {journal} {Rep. Prog. Phys.}\ }\textbf {\bibinfo {volume} {62}},\ \bibinfo
  {pages} {61} (\bibinfo {year} {1999})}\BibitemShut {NoStop}%
\bibitem [{\citenamefont {Keimer}\ \emph {et~al.}(2015)\citenamefont {Keimer},
  \citenamefont {Kivelson}, \citenamefont {Norman}, \citenamefont {Uchida},\
  and\ \citenamefont {Zaanen}}]{Keimer15}%
  \BibitemOpen
  \bibfield  {author} {\bibinfo {author} {\bibfnamefont {B.}~\bibnamefont
  {Keimer}}, \bibinfo {author} {\bibfnamefont {S.~A.}\ \bibnamefont
  {Kivelson}}, \bibinfo {author} {\bibfnamefont {M.~R.}\ \bibnamefont
  {Norman}}, \bibinfo {author} {\bibfnamefont {S.}~\bibnamefont {Uchida}},\
  and\ \bibinfo {author} {\bibfnamefont {J.}~\bibnamefont {Zaanen}},\
  }\bibfield  {title} {\bibinfo {title} {From quantum matter to
  high-temperature superconductivity in copper oxides},\ }\href@noop {}
  {\bibfield  {journal} {\bibinfo  {journal} {Nature}\ }\textbf {\bibinfo
  {volume} {518}},\ \bibinfo {pages} {179} (\bibinfo {year}
  {2015})}\BibitemShut {NoStop}%
\bibitem [{\citenamefont {L\"ohneysen}\ \emph {et~al.}(2007)\citenamefont
  {L\"ohneysen}, \citenamefont {Rosch}, \citenamefont {Vojta},\ and\
  \citenamefont {W\"olfle}}]{Lohneysen07}%
  \BibitemOpen
  \bibfield  {author} {\bibinfo {author} {\bibfnamefont {H.~v.}\ \bibnamefont
  {L\"ohneysen}}, \bibinfo {author} {\bibfnamefont {A.}~\bibnamefont {Rosch}},
  \bibinfo {author} {\bibfnamefont {M.}~\bibnamefont {Vojta}},\ and\ \bibinfo
  {author} {\bibfnamefont {P.}~\bibnamefont {W\"olfle}},\ }\bibfield  {title}
  {\bibinfo {title} {Fermi-liquid instabilities at magnetic quantum phase
  transitions},\ }\href {https://doi.org/10.1103/RevModPhys.79.1015} {\bibfield
   {journal} {\bibinfo  {journal} {Rev. Mod. Phys.}\ }\textbf {\bibinfo
  {volume} {79}},\ \bibinfo {pages} {1015} (\bibinfo {year}
  {2007})}\BibitemShut {NoStop}%
\bibitem [{\citenamefont {Gegenwart}\ \emph {et~al.}(2008)\citenamefont
  {Gegenwart}, \citenamefont {Si},\ and\ \citenamefont
  {Steglich}}]{Gegenwart08}%
  \BibitemOpen
  \bibfield  {author} {\bibinfo {author} {\bibfnamefont {P.}~\bibnamefont
  {Gegenwart}}, \bibinfo {author} {\bibfnamefont {Q.}~\bibnamefont {Si}},\ and\
  \bibinfo {author} {\bibfnamefont {F.}~\bibnamefont {Steglich}},\ }\bibfield
  {title} {\bibinfo {title} {Quantum criticality in heavy-fermion metals},\
  }\href@noop {} {\bibfield  {journal} {\bibinfo  {journal} {Nat. Phys.}\
  }\textbf {\bibinfo {volume} {4}},\ \bibinfo {pages} {186} (\bibinfo {year}
  {2008})}\BibitemShut {NoStop}%
\bibitem [{\citenamefont {Shibauchi}\ \emph {et~al.}(2014)\citenamefont
  {Shibauchi}, \citenamefont {Carrington},\ and\ \citenamefont
  {Matsuda}}]{Shibauchi14}%
  \BibitemOpen
  \bibfield  {author} {\bibinfo {author} {\bibfnamefont {T.}~\bibnamefont
  {Shibauchi}}, \bibinfo {author} {\bibfnamefont {A.}~\bibnamefont
  {Carrington}},\ and\ \bibinfo {author} {\bibfnamefont {Y.}~\bibnamefont
  {Matsuda}},\ }\bibfield  {title} {\bibinfo {title} {A quantum critical point
  lying beneath the superconducting dome in iron pnictides},\ }\href
  {https://doi.org/10.1146/annurev-conmatphys-031113-133921} {\bibfield
  {journal} {\bibinfo  {journal} {Ann. Rev. Cond. Mat. Phys.}\ }\textbf
  {\bibinfo {volume} {5}},\ \bibinfo {pages} {113} (\bibinfo {year}
  {2014})}\BibitemShut {NoStop}%
\bibitem [{\citenamefont {Zhang}\ \emph {et~al.}(2012)\citenamefont {Zhang},
  \citenamefont {Hung}, \citenamefont {Tung},\ and\ \citenamefont
  {Chin}}]{Zhang12}%
  \BibitemOpen
  \bibfield  {author} {\bibinfo {author} {\bibfnamefont {X.}~\bibnamefont
  {Zhang}}, \bibinfo {author} {\bibfnamefont {C.-L.}\ \bibnamefont {Hung}},
  \bibinfo {author} {\bibfnamefont {S.-K.}\ \bibnamefont {Tung}},\ and\
  \bibinfo {author} {\bibfnamefont {C.}~\bibnamefont {Chin}},\ }\bibfield
  {title} {\bibinfo {title} {Observation of quantum criticality with ultracold
  atoms in optical lattices},\ }\href {https://doi.org/10.1126/science.1217990}
  {\bibfield  {journal} {\bibinfo  {journal} {Science}\ }\textbf {\bibinfo
  {volume} {335}},\ \bibinfo {pages} {1070} (\bibinfo {year}
  {2012})}\BibitemShut {NoStop}%
\bibitem [{\citenamefont {Coldea}\ \emph {et~al.}(2010)\citenamefont {Coldea},
  \citenamefont {Tennant}, \citenamefont {Wheeler}, \citenamefont {Wawrzynska},
  \citenamefont {Prabhakaran}, \citenamefont {Telling}, \citenamefont
  {Habicht}, \citenamefont {Smeibidl},\ and\ \citenamefont
  {Kiefer}}]{Coldea10}%
  \BibitemOpen
  \bibfield  {author} {\bibinfo {author} {\bibfnamefont {R.}~\bibnamefont
  {Coldea}}, \bibinfo {author} {\bibfnamefont {D.~A.}\ \bibnamefont {Tennant}},
  \bibinfo {author} {\bibfnamefont {E.~M.}\ \bibnamefont {Wheeler}}, \bibinfo
  {author} {\bibfnamefont {E.}~\bibnamefont {Wawrzynska}}, \bibinfo {author}
  {\bibfnamefont {D.}~\bibnamefont {Prabhakaran}}, \bibinfo {author}
  {\bibfnamefont {M.}~\bibnamefont {Telling}}, \bibinfo {author} {\bibfnamefont
  {K.}~\bibnamefont {Habicht}}, \bibinfo {author} {\bibfnamefont
  {P.}~\bibnamefont {Smeibidl}},\ and\ \bibinfo {author} {\bibfnamefont
  {K.}~\bibnamefont {Kiefer}},\ }\bibfield  {title} {\bibinfo {title} {Quantum
  criticality in an ising chain: Experimental evidence for emergent
  $\mathrm{E}_{\mathrm{8}}$ symmetry},\ }\href
  {https://doi.org/10.1126/science.1180085} {\bibfield  {journal} {\bibinfo
  {journal} {Science}\ }\textbf {\bibinfo {volume} {327}},\ \bibinfo {pages}
  {177} (\bibinfo {year} {2010})}\BibitemShut {NoStop}%
\bibitem [{\citenamefont {Paglione}\ \emph {et~al.}(2003)\citenamefont
  {Paglione}, \citenamefont {Tanatar}, \citenamefont {Hawthorn}, \citenamefont
  {Boaknin}, \citenamefont {Hill}, \citenamefont {Ronning}, \citenamefont
  {Sutherland}, \citenamefont {Taillefer}, \citenamefont {Petrovic},\ and\
  \citenamefont {Canfield}}]{Paglione03}%
  \BibitemOpen
  \bibfield  {author} {\bibinfo {author} {\bibfnamefont {J.}~\bibnamefont
  {Paglione}}, \bibinfo {author} {\bibfnamefont {M.~A.}\ \bibnamefont
  {Tanatar}}, \bibinfo {author} {\bibfnamefont {D.~G.}\ \bibnamefont
  {Hawthorn}}, \bibinfo {author} {\bibfnamefont {E.}~\bibnamefont {Boaknin}},
  \bibinfo {author} {\bibfnamefont {R.~W.}\ \bibnamefont {Hill}}, \bibinfo
  {author} {\bibfnamefont {F.}~\bibnamefont {Ronning}}, \bibinfo {author}
  {\bibfnamefont {M.}~\bibnamefont {Sutherland}}, \bibinfo {author}
  {\bibfnamefont {L.}~\bibnamefont {Taillefer}}, \bibinfo {author}
  {\bibfnamefont {C.}~\bibnamefont {Petrovic}},\ and\ \bibinfo {author}
  {\bibfnamefont {P.~C.}\ \bibnamefont {Canfield}},\ }\bibfield  {title}
  {\bibinfo {title} {Field-induced quantum critical point in
  $\mathrm{C}{\mathrm{e}}\mathrm{C}{\mathrm{o}}\mathrm{I}{\mathrm{n}}_{\mathrm{5}}$},\
  }\href {https://doi.org/10.1103/PhysRevLett.91.246405} {\bibfield  {journal}
  {\bibinfo  {journal} {Phys. Rev. Lett.}\ }\textbf {\bibinfo {volume} {91}},\
  \bibinfo {pages} {246405} (\bibinfo {year} {2003})}\BibitemShut {NoStop}%
\bibitem [{\citenamefont {Oosawa}\ \emph {et~al.}(2003)\citenamefont {Oosawa},
  \citenamefont {Fujisawa}, \citenamefont {Osakabe}, \citenamefont {Kakurai},\
  and\ \citenamefont {Tanaka}}]{Oosawa03}%
  \BibitemOpen
  \bibfield  {author} {\bibinfo {author} {\bibfnamefont {A.}~\bibnamefont
  {Oosawa}}, \bibinfo {author} {\bibfnamefont {M.}~\bibnamefont {Fujisawa}},
  \bibinfo {author} {\bibfnamefont {T.}~\bibnamefont {Osakabe}}, \bibinfo
  {author} {\bibfnamefont {K.}~\bibnamefont {Kakurai}},\ and\ \bibinfo {author}
  {\bibfnamefont {H.}~\bibnamefont {Tanaka}},\ }\bibfield  {title} {\bibinfo
  {title} {Neutron diffraction study of the pressure-induced magnetic ordering
  in the spin gap system
  $\mathrm{T}{\mathrm{l}}\mathrm{C}{\mathrm{u}}\mathrm{C}{\mathrm{l}}_{\mathrm{3}}$},\
  }\href@noop {} {\bibfield  {journal} {\bibinfo  {journal} {J. Phys. Soc.
  Jpn.}\ }\textbf {\bibinfo {volume} {72}},\ \bibinfo {pages} {1026} (\bibinfo
  {year} {2003})}\BibitemShut {NoStop}%
\bibitem [{\citenamefont {R{\"u}egg}\ \emph {et~al.}(2003)\citenamefont
  {R{\"u}egg}, \citenamefont {Cavadini}, \citenamefont {Furrer}, \citenamefont
  {G{\"u}del}, \citenamefont {Kr{\"a}mer}, \citenamefont {Mutka}, \citenamefont
  {Wildes}, \citenamefont {Habicht},\ and\ \citenamefont
  {Vorderwisch}}]{Ruegg03}%
  \BibitemOpen
  \bibfield  {author} {\bibinfo {author} {\bibfnamefont {C.}~\bibnamefont
  {R{\"u}egg}}, \bibinfo {author} {\bibfnamefont {N.}~\bibnamefont {Cavadini}},
  \bibinfo {author} {\bibfnamefont {A.}~\bibnamefont {Furrer}}, \bibinfo
  {author} {\bibfnamefont {H.-U.}\ \bibnamefont {G{\"u}del}}, \bibinfo {author}
  {\bibfnamefont {K.}~\bibnamefont {Kr{\"a}mer}}, \bibinfo {author}
  {\bibfnamefont {H.}~\bibnamefont {Mutka}}, \bibinfo {author} {\bibfnamefont
  {A.}~\bibnamefont {Wildes}}, \bibinfo {author} {\bibfnamefont
  {K.}~\bibnamefont {Habicht}},\ and\ \bibinfo {author} {\bibfnamefont
  {P.}~\bibnamefont {Vorderwisch}},\ }\bibfield  {title} {\bibinfo {title}
  {Bose--einstein condensation of the triplet states in the magnetic insulator
  $\mathrm{T}{\mathrm{l}}\mathrm{C}{\mathrm{u}}\mathrm{C}{\mathrm{l}}_{\mathrm{3}}$},\
  }\href@noop {} {\bibfield  {journal} {\bibinfo  {journal} {Nature}\ }\textbf
  {\bibinfo {volume} {423}},\ \bibinfo {pages} {62} (\bibinfo {year}
  {2003})}\BibitemShut {NoStop}%
\bibitem [{\citenamefont {Tanaka}\ \emph {et~al.}(2003)\citenamefont {Tanaka},
  \citenamefont {Goto}, \citenamefont {Fujisawa}, \citenamefont {Ono},\ and\
  \citenamefont {Uwatoko}}]{Tanaka03}%
  \BibitemOpen
  \bibfield  {author} {\bibinfo {author} {\bibfnamefont {H.}~\bibnamefont
  {Tanaka}}, \bibinfo {author} {\bibfnamefont {K.}~\bibnamefont {Goto}},
  \bibinfo {author} {\bibfnamefont {M.}~\bibnamefont {Fujisawa}}, \bibinfo
  {author} {\bibfnamefont {T.}~\bibnamefont {Ono}},\ and\ \bibinfo {author}
  {\bibfnamefont {Y.}~\bibnamefont {Uwatoko}},\ }\bibfield  {title} {\bibinfo
  {title} {Magnetic ordering under high pressure in the quantum spin system
  $\mathrm{T}{\mathrm{l}}\mathrm{C}{\mathrm{u}}\mathrm{C}{\mathrm{l}}_{\mathrm{3}}$},\
  }\href {https://doi.org/https://doi.org/10.1016/S0921-4526(02)02009-4}
  {\bibfield  {journal} {\bibinfo  {journal} {Physica B: Condensed Matter}\
  }\textbf {\bibinfo {volume} {329-333}},\ \bibinfo {pages} {697 } (\bibinfo
  {year} {2003})}\BibitemShut {NoStop}%
\bibitem [{\citenamefont {R\"uegg}\ \emph {et~al.}(2004)\citenamefont
  {R\"uegg}, \citenamefont {Furrer}, \citenamefont {Sheptyakov}, \citenamefont
  {Str\"assle}, \citenamefont {Kr\"amer}, \citenamefont {G\"udel},\ and\
  \citenamefont {M\'el\'esi}}]{Ruegg04}%
  \BibitemOpen
  \bibfield  {author} {\bibinfo {author} {\bibfnamefont {C.}~\bibnamefont
  {R\"uegg}}, \bibinfo {author} {\bibfnamefont {A.}~\bibnamefont {Furrer}},
  \bibinfo {author} {\bibfnamefont {D.}~\bibnamefont {Sheptyakov}}, \bibinfo
  {author} {\bibfnamefont {T.}~\bibnamefont {Str\"assle}}, \bibinfo {author}
  {\bibfnamefont {K.~W.}\ \bibnamefont {Kr\"amer}}, \bibinfo {author}
  {\bibfnamefont {H.-U.}\ \bibnamefont {G\"udel}},\ and\ \bibinfo {author}
  {\bibfnamefont {L.}~\bibnamefont {M\'el\'esi}},\ }\bibfield  {title}
  {\bibinfo {title} {Pressure-induced quantum phase transition in the
  spin-liquid
  $\mathrm{T}\mathrm{l}\mathrm{C}\mathrm{u}{\mathrm{c}\mathrm{l}}_{3}$},\
  }\href {https://doi.org/10.1103/PhysRevLett.93.257201} {\bibfield  {journal}
  {\bibinfo  {journal} {Phys. Rev. Lett.}\ }\textbf {\bibinfo {volume} {93}},\
  \bibinfo {pages} {257201} (\bibinfo {year} {2004})}\BibitemShut {NoStop}%
\bibitem [{\citenamefont {R\"uegg}\ \emph {et~al.}(2008)\citenamefont
  {R\"uegg}, \citenamefont {Normand}, \citenamefont {Matsumoto}, \citenamefont
  {Furrer}, \citenamefont {McMorrow}, \citenamefont {Kr\"amer}, \citenamefont
  {G\"udel}, \citenamefont {Gvasaliya}, \citenamefont {Mutka},\ and\
  \citenamefont {Boehm}}]{Ruegg08}%
  \BibitemOpen
  \bibfield  {author} {\bibinfo {author} {\bibfnamefont {C.}~\bibnamefont
  {R\"uegg}}, \bibinfo {author} {\bibfnamefont {B.}~\bibnamefont {Normand}},
  \bibinfo {author} {\bibfnamefont {M.}~\bibnamefont {Matsumoto}}, \bibinfo
  {author} {\bibfnamefont {A.}~\bibnamefont {Furrer}}, \bibinfo {author}
  {\bibfnamefont {D.~F.}\ \bibnamefont {McMorrow}}, \bibinfo {author}
  {\bibfnamefont {K.~W.}\ \bibnamefont {Kr\"amer}}, \bibinfo {author}
  {\bibfnamefont {H.~U.}\ \bibnamefont {G\"udel}}, \bibinfo {author}
  {\bibfnamefont {S.~N.}\ \bibnamefont {Gvasaliya}}, \bibinfo {author}
  {\bibfnamefont {H.}~\bibnamefont {Mutka}},\ and\ \bibinfo {author}
  {\bibfnamefont {M.}~\bibnamefont {Boehm}},\ }\bibfield  {title} {\bibinfo
  {title} {Quantum magnets under pressure: Controlling elementary excitations
  in
  $\mathrm{T}{\mathrm{l}}\mathrm{C}{\mathrm{u}}\mathrm{C}{\mathrm{l}}_{\mathrm{3}}$},\
  }\href {https://doi.org/10.1103/PhysRevLett.100.205701} {\bibfield  {journal}
  {\bibinfo  {journal} {Phys. Rev. Lett.}\ }\textbf {\bibinfo {volume} {100}},\
  \bibinfo {pages} {205701} (\bibinfo {year} {2008})}\BibitemShut {NoStop}%
\bibitem [{\citenamefont {Hase}\ \emph {et~al.}(2016)\citenamefont {Hase},
  \citenamefont {Matsumoto}, \citenamefont {Matsuo},\ and\ \citenamefont
  {Kindo}}]{Hase16}%
  \BibitemOpen
  \bibfield  {author} {\bibinfo {author} {\bibfnamefont {M.}~\bibnamefont
  {Hase}}, \bibinfo {author} {\bibfnamefont {M.}~\bibnamefont {Matsumoto}},
  \bibinfo {author} {\bibfnamefont {A.}~\bibnamefont {Matsuo}},\ and\ \bibinfo
  {author} {\bibfnamefont {K.}~\bibnamefont {Kindo}},\ }\bibfield  {title}
  {\bibinfo {title} {Magnetism of the antiferromagnetic spin-$\frac{1}{2}$
  tetramer compound
  $\mathrm{C}{\mathrm{u}}\mathrm{I}{\mathrm{n}}\mathrm{V}\mathrm{O}_{\mathrm{5}}$},\
  }\href {https://doi.org/10.1103/PhysRevB.94.174421} {\bibfield  {journal}
  {\bibinfo  {journal} {Phys. Rev. B}\ }\textbf {\bibinfo {volume} {94}},\
  \bibinfo {pages} {174421} (\bibinfo {year} {2016})}\BibitemShut {NoStop}%
\bibitem [{\citenamefont {Singhania}\ and\ \citenamefont
  {Kumar}(2018)}]{Singhania18}%
  \BibitemOpen
  \bibfield  {author} {\bibinfo {author} {\bibfnamefont {A.}~\bibnamefont
  {Singhania}}\ and\ \bibinfo {author} {\bibfnamefont {S.}~\bibnamefont
  {Kumar}},\ }\bibfield  {title} {\bibinfo {title} {Cluster mean-field study of
  the heisenberg model for
  $\mathrm{C}{\mathrm{u}}\mathrm{I}{\mathrm{n}}\mathrm{V}\mathrm{O}_{\mathrm{5}}$},\
  }\href {https://doi.org/10.1103/PhysRevB.98.104429} {\bibfield  {journal}
  {\bibinfo  {journal} {Phys. Rev. B}\ }\textbf {\bibinfo {volume} {98}},\
  \bibinfo {pages} {104429} (\bibinfo {year} {2018})}\BibitemShut {NoStop}%
\bibitem [{\citenamefont {Moser}\ \emph {et~al.}(1999)\citenamefont {Moser},
  \citenamefont {Cirpus},\ and\ \citenamefont {Jung}}]{Moser99}%
  \BibitemOpen
  \bibfield  {author} {\bibinfo {author} {\bibfnamefont {P.}~\bibnamefont
  {Moser}}, \bibinfo {author} {\bibfnamefont {V.}~\bibnamefont {Cirpus}},\ and\
  \bibinfo {author} {\bibfnamefont {W.}~\bibnamefont {Jung}},\ }\bibfield
  {title} {\bibinfo {title}
  {$\mathrm{C}{\mathrm{u}}\mathrm{I}{\mathrm{n}}\mathrm{O}\mathrm{V}\mathrm{O}_{\mathrm{4}}$
  – $\mathrm{E}$inkristalle eines $\mathrm{K}$upfer(ii)-indiumoxidvanadats
  durch oxidation von
  $\mathrm{C}{\mathrm{u}}$/$\mathrm{I}{\mathrm{n}}$/$\mathrm{V}$-legierungen},\
  }\href
  {https://doi.org/10.1002/(SICI)1521-3749(199905)625:5<714::AID-ZAAC714>3.0.CO;2-0}
  {\bibfield  {journal} {\bibinfo  {journal} {Zeitschrift f\"ur anorganische
  und allgemeine Chemie}\ }\textbf {\bibinfo {volume} {625}},\ \bibinfo {pages}
  {714} (\bibinfo {year} {1999})}\BibitemShut {NoStop}%
\bibitem [{\citenamefont {Carmalt}\ \emph {et~al.}(1995)\citenamefont
  {Carmalt}, \citenamefont {Farrugia},\ and\ \citenamefont
  {Norman}}]{Carmalt95}%
  \BibitemOpen
  \bibfield  {author} {\bibinfo {author} {\bibfnamefont {C.~J.}\ \bibnamefont
  {Carmalt}}, \bibinfo {author} {\bibfnamefont {L.~J.}\ \bibnamefont
  {Farrugia}},\ and\ \bibinfo {author} {\bibfnamefont {N.~C.}\ \bibnamefont
  {Norman}},\ }\bibfield  {title} {\bibinfo {title} {Synthesis and
  $\mathrm{X}$-ray crystal structure of a polymeric iodobismuthate anion},\
  }\href@noop {} {\bibfield  {journal} {\bibinfo  {journal} {Zeitschrift
  f{\"u}r Naturforschung B}\ }\textbf {\bibinfo {volume} {50}},\ \bibinfo
  {pages} {1591} (\bibinfo {year} {1995})}\BibitemShut {NoStop}%
\bibitem [{\citenamefont {Hase}\ \emph {et~al.}(2015)\citenamefont {Hase},
  \citenamefont {Nakajima}, \citenamefont {Ohira-Kawamura}, \citenamefont
  {Kawakita}, \citenamefont {Kikuchi},\ and\ \citenamefont
  {Matsumoto}}]{Hase15}%
  \BibitemOpen
  \bibfield  {author} {\bibinfo {author} {\bibfnamefont {M.}~\bibnamefont
  {Hase}}, \bibinfo {author} {\bibfnamefont {K.}~\bibnamefont {Nakajima}},
  \bibinfo {author} {\bibfnamefont {S.}~\bibnamefont {Ohira-Kawamura}},
  \bibinfo {author} {\bibfnamefont {Y.}~\bibnamefont {Kawakita}}, \bibinfo
  {author} {\bibfnamefont {T.}~\bibnamefont {Kikuchi}},\ and\ \bibinfo {author}
  {\bibfnamefont {M.}~\bibnamefont {Matsumoto}},\ }\bibfield  {title} {\bibinfo
  {title} {Magnetic excitations in the spin-$\frac{1}{2}$ tetramer substance
  $\mathrm{Cu}_{2}^{114}\mathrm{C}{\mathrm{d}}^{11}\mathrm{B}_{2}\mathrm{O}_{6}$
  obtained by inelastic neutron scattering experiments},\ }\href
  {https://doi.org/10.1103/PhysRevB.92.184412} {\bibfield  {journal} {\bibinfo
  {journal} {Phys. Rev. B}\ }\textbf {\bibinfo {volume} {92}},\ \bibinfo
  {pages} {184412} (\bibinfo {year} {2015})}\BibitemShut {NoStop}%
\bibitem [{\citenamefont {Effenberger}(1986)}]{Effenberger86}%
  \BibitemOpen
  \bibfield  {author} {\bibinfo {author} {\bibfnamefont {H.}~\bibnamefont
  {Effenberger}},\ }\bibfield  {title} {\bibinfo {title} {Die
  kristallstrukturen von drei modifikationen des
  $\mathrm{C}{\mathrm{u}}(\mathrm{S}{\mathrm{e}}\mathrm{O}_{3})$},\ }\href@noop
  {} {\bibfield  {journal} {\bibinfo  {journal} {Zeitschrift f{\"u}r
  Kristallographie-Crystalline Materials}\ }\textbf {\bibinfo {volume} {175}},\
  \bibinfo {pages} {61} (\bibinfo {year} {1986})}\BibitemShut {NoStop}%
\bibitem [{\citenamefont {\ifmmode \check{Z}\else
  \v{Z}\fi{}ivkovi\ifmmode~\acute{c}\else \'{c}\fi{}}\ \emph
  {et~al.}(2012)\citenamefont {\ifmmode \check{Z}\else
  \v{Z}\fi{}ivkovi\ifmmode~\acute{c}\else \'{c}\fi{}}, \citenamefont
  {Djoki\ifmmode~\acute{c}\else \'{c}\fi{}}, \citenamefont {Herak},
  \citenamefont {Paji\ifmmode~\acute{c}\else \'{c}\fi{}}, \citenamefont
  {Pr\ifmmode~\check{s}\else \v{s}\fi{}a}, \citenamefont {Pattison},
  \citenamefont {Dominko}, \citenamefont {Mickovi\ifmmode~\acute{c}\else
  \'{c}\fi{}}, \citenamefont {Cin\ifmmode \check{c}\else
  \v{c}\fi{}i\ifmmode~\acute{c}\else \'{c}\fi{}}, \citenamefont {Forr\'o},
  \citenamefont {Berger},\ and\ \citenamefont {R\o{}nnow}}]{Zivkovic12}%
  \BibitemOpen
  \bibfield  {author} {\bibinfo {author} {\bibfnamefont {I.}~\bibnamefont
  {\ifmmode \check{Z}\else \v{Z}\fi{}ivkovi\ifmmode~\acute{c}\else
  \'{c}\fi{}}}, \bibinfo {author} {\bibfnamefont {D.~M.}\ \bibnamefont
  {Djoki\ifmmode~\acute{c}\else \'{c}\fi{}}}, \bibinfo {author} {\bibfnamefont
  {M.}~\bibnamefont {Herak}}, \bibinfo {author} {\bibfnamefont
  {D.}~\bibnamefont {Paji\ifmmode~\acute{c}\else \'{c}\fi{}}}, \bibinfo
  {author} {\bibfnamefont {K.}~\bibnamefont {Pr\ifmmode~\check{s}\else
  \v{s}\fi{}a}}, \bibinfo {author} {\bibfnamefont {P.}~\bibnamefont
  {Pattison}}, \bibinfo {author} {\bibfnamefont {D.}~\bibnamefont {Dominko}},
  \bibinfo {author} {\bibfnamefont {Z.}~\bibnamefont
  {Mickovi\ifmmode~\acute{c}\else \'{c}\fi{}}}, \bibinfo {author}
  {\bibfnamefont {D.}~\bibnamefont {Cin\ifmmode \check{c}\else
  \v{c}\fi{}i\ifmmode~\acute{c}\else \'{c}\fi{}}}, \bibinfo {author}
  {\bibfnamefont {L.}~\bibnamefont {Forr\'o}}, \bibinfo {author} {\bibfnamefont
  {H.}~\bibnamefont {Berger}},\ and\ \bibinfo {author} {\bibfnamefont {H.~M.}\
  \bibnamefont {R\o{}nnow}},\ }\bibfield  {title} {\bibinfo {title}
  {Site-selective quantum correlations revealed by magnetic anisotropy in the
  tetramer system secuo${}_{3}$},\ }\href
  {https://doi.org/10.1103/PhysRevB.86.054405} {\bibfield  {journal} {\bibinfo
  {journal} {Phys. Rev. B}\ }\textbf {\bibinfo {volume} {86}},\ \bibinfo
  {pages} {054405} (\bibinfo {year} {2012})}\BibitemShut {NoStop}%
\bibitem [{\citenamefont {Matsumoto}\ \emph {et~al.}(2010)\citenamefont
  {Matsumoto}, \citenamefont {Kuroe}, \citenamefont {Sekine},\ and\
  \citenamefont {Masuda}}]{Matsumoto10}%
  \BibitemOpen
  \bibfield  {author} {\bibinfo {author} {\bibfnamefont {M.}~\bibnamefont
  {Matsumoto}}, \bibinfo {author} {\bibfnamefont {H.}~\bibnamefont {Kuroe}},
  \bibinfo {author} {\bibfnamefont {T.}~\bibnamefont {Sekine}},\ and\ \bibinfo
  {author} {\bibfnamefont {T.}~\bibnamefont {Masuda}},\ }\bibfield  {title}
  {\bibinfo {title} {Transverse and longitudinal excitation modes in
  interacting multispin systems},\ }\href
  {https://doi.org/10.1143/JPSJ.79.084703} {\bibfield  {journal} {\bibinfo
  {journal} {J. Phys. Soc. Jpn.}\ }\textbf {\bibinfo {volume} {79}},\ \bibinfo
  {pages} {084703} (\bibinfo {year} {2010})},\ \Eprint
  {https://arxiv.org/abs/https://doi.org/10.1143/JPSJ.79.084703}
  {https://doi.org/10.1143/JPSJ.79.084703} \BibitemShut {NoStop}%
\bibitem [{\citenamefont {Masuda}\ \emph {et~al.}(2004)\citenamefont {Masuda},
  \citenamefont {Zheludev}, \citenamefont {Grenier}, \citenamefont {Imai},
  \citenamefont {Uchinokura}, \citenamefont {Ressouche},\ and\ \citenamefont
  {Park}}]{Masuda04}%
  \BibitemOpen
  \bibfield  {author} {\bibinfo {author} {\bibfnamefont {T.}~\bibnamefont
  {Masuda}}, \bibinfo {author} {\bibfnamefont {A.}~\bibnamefont {Zheludev}},
  \bibinfo {author} {\bibfnamefont {B.}~\bibnamefont {Grenier}}, \bibinfo
  {author} {\bibfnamefont {S.}~\bibnamefont {Imai}}, \bibinfo {author}
  {\bibfnamefont {K.}~\bibnamefont {Uchinokura}}, \bibinfo {author}
  {\bibfnamefont {E.}~\bibnamefont {Ressouche}},\ and\ \bibinfo {author}
  {\bibfnamefont {S.}~\bibnamefont {Park}},\ }\bibfield  {title} {\bibinfo
  {title} {Cooperative ordering of gapped and gapless spin networks in
  $\mathrm{C}{\mathrm{u}}_{\mathrm{2}}$$\mathrm{F}{\mathrm{e}}_{\mathrm{2}}\mathrm{G}{\mathrm{e}}_{\mathrm{4}}\mathrm{O}_{\mathrm{13}}$},\
  }\href {https://doi.org/10.1103/PhysRevLett.93.077202} {\bibfield  {journal}
  {\bibinfo  {journal} {Phys. Rev. Lett.}\ }\textbf {\bibinfo {volume} {93}},\
  \bibinfo {pages} {077202} (\bibinfo {year} {2004})}\BibitemShut {NoStop}%
\bibitem [{\citenamefont {Klevtsova}\ and\ \citenamefont
  {Glinskaya}(1982)}]{Klevtsova82}%
  \BibitemOpen
  \bibfield  {author} {\bibinfo {author} {\bibfnamefont {R.~F.}\ \bibnamefont
  {Klevtsova}}\ and\ \bibinfo {author} {\bibfnamefont {L.~A.}\ \bibnamefont
  {Glinskaya}},\ }\bibfield  {title} {\bibinfo {title} {Crystal structure of
  potassium nickel molybdate
  $\mathrm{K}_{2}\mathrm{N}{\mathrm{i}}_{2}(\mathrm{M}{\mathrm{o}}\mathrm{O}_{4})_{3}$},\
  }\href {https://doi.org/10.1007/BF00746215} {\bibfield  {journal} {\bibinfo
  {journal} {Journal of Structural Chemistry}\ }\textbf {\bibinfo {volume}
  {23}},\ \bibinfo {pages} {816} (\bibinfo {year} {1982})}\BibitemShut
  {NoStop}%
\bibitem [{\citenamefont {Hase}\ \emph {et~al.}(2017)\citenamefont {Hase},
  \citenamefont {Matsuo}, \citenamefont {Kindo},\ and\ \citenamefont
  {Matsumoto}}]{Hase17}%
  \BibitemOpen
  \bibfield  {author} {\bibinfo {author} {\bibfnamefont {M.}~\bibnamefont
  {Hase}}, \bibinfo {author} {\bibfnamefont {A.}~\bibnamefont {Matsuo}},
  \bibinfo {author} {\bibfnamefont {K.}~\bibnamefont {Kindo}},\ and\ \bibinfo
  {author} {\bibfnamefont {M.}~\bibnamefont {Matsumoto}},\ }\bibfield  {title}
  {\bibinfo {title} {Magnetism of the spin-1 tetramer compound
  ${A}_{2}\mathrm{N}{\mathrm{i}}_{2}\mathrm{M}{\mathrm{o}}_{3}\mathrm{O}_{12}
  ({A}=\text{Rb or K})$},\ }\href {https://doi.org/10.1103/PhysRevB.96.214424}
  {\bibfield  {journal} {\bibinfo  {journal} {Phys. Rev. B}\ }\textbf {\bibinfo
  {volume} {96}},\ \bibinfo {pages} {214424} (\bibinfo {year}
  {2017})}\BibitemShut {NoStop}%
\bibitem [{\citenamefont {White}(1992)}]{White92}%
  \BibitemOpen
  \bibfield  {author} {\bibinfo {author} {\bibfnamefont {S.~R.}\ \bibnamefont
  {White}},\ }\bibfield  {title} {\bibinfo {title} {Density matrix formulation
  for quantum renormalization groups},\ }\href
  {https://doi.org/10.1103/PhysRevLett.69.2863} {\bibfield  {journal} {\bibinfo
   {journal} {Phys. Rev. Lett.}\ }\textbf {\bibinfo {volume} {69}},\ \bibinfo
  {pages} {2863} (\bibinfo {year} {1992})}\BibitemShut {NoStop}%
\bibitem [{\citenamefont {Yasuda}\ \emph {et~al.}(2005)\citenamefont {Yasuda},
  \citenamefont {Todo}, \citenamefont {Hukushima}, \citenamefont {Alet},
  \citenamefont {Keller}, \citenamefont {Troyer},\ and\ \citenamefont
  {Takayama}}]{Yasuda05}%
  \BibitemOpen
  \bibfield  {author} {\bibinfo {author} {\bibfnamefont {C.}~\bibnamefont
  {Yasuda}}, \bibinfo {author} {\bibfnamefont {S.}~\bibnamefont {Todo}},
  \bibinfo {author} {\bibfnamefont {K.}~\bibnamefont {Hukushima}}, \bibinfo
  {author} {\bibfnamefont {F.}~\bibnamefont {Alet}}, \bibinfo {author}
  {\bibfnamefont {M.}~\bibnamefont {Keller}}, \bibinfo {author} {\bibfnamefont
  {M.}~\bibnamefont {Troyer}},\ and\ \bibinfo {author} {\bibfnamefont
  {H.}~\bibnamefont {Takayama}},\ }\bibfield  {title} {\bibinfo {title} {N\'eel
  temperature of quasi-low-dimensional heisenberg antiferromagnets},\ }\href
  {https://doi.org/10.1103/PhysRevLett.94.217201} {\bibfield  {journal}
  {\bibinfo  {journal} {Phys. Rev. Lett.}\ }\textbf {\bibinfo {volume} {94}},\
  \bibinfo {pages} {217201} (\bibinfo {year} {2005})}\BibitemShut {NoStop}%
\bibitem [{\citenamefont {Affleck}\ \emph {et~al.}(1987)\citenamefont
  {Affleck}, \citenamefont {Kennedy}, \citenamefont {Lieb},\ and\ \citenamefont
  {Tasaki}}]{Affleck87}%
  \BibitemOpen
  \bibfield  {author} {\bibinfo {author} {\bibfnamefont {I.}~\bibnamefont
  {Affleck}}, \bibinfo {author} {\bibfnamefont {T.}~\bibnamefont {Kennedy}},
  \bibinfo {author} {\bibfnamefont {E.~H.}\ \bibnamefont {Lieb}},\ and\
  \bibinfo {author} {\bibfnamefont {H.}~\bibnamefont {Tasaki}},\ }\bibfield
  {title} {\bibinfo {title} {Rigorous results on valence-bond ground states in
  antiferromagnets},\ }\href {https://doi.org/10.1103/PhysRevLett.59.799}
  {\bibfield  {journal} {\bibinfo  {journal} {Phys. Rev. Lett.}\ }\textbf
  {\bibinfo {volume} {59}},\ \bibinfo {pages} {799} (\bibinfo {year}
  {1987})}\BibitemShut {NoStop}%
\bibitem [{\citenamefont {den Nijs}\ and\ \citenamefont
  {Rommelse}(1989)}]{deNijs89}%
  \BibitemOpen
  \bibfield  {author} {\bibinfo {author} {\bibfnamefont {M.}~\bibnamefont {den
  Nijs}}\ and\ \bibinfo {author} {\bibfnamefont {K.}~\bibnamefont {Rommelse}},\
  }\bibfield  {title} {\bibinfo {title} {Preroughening transitions in crystal
  surfaces and valence-bond phases in quantum spin chains},\ }\href
  {https://doi.org/10.1103/PhysRevB.40.4709} {\bibfield  {journal} {\bibinfo
  {journal} {Phys. Rev. B}\ }\textbf {\bibinfo {volume} {40}},\ \bibinfo
  {pages} {4709} (\bibinfo {year} {1989})}\BibitemShut {NoStop}%
\bibitem [{\citenamefont {Agrapidis}\ \emph {et~al.}(2019)\citenamefont
  {Agrapidis}, \citenamefont {Drechsler}, \citenamefont {van~den Brink},\ and\
  \citenamefont {Nishimoto}}]{Agrapidis18}%
  \BibitemOpen
  \bibfield  {author} {\bibinfo {author} {\bibfnamefont {C.~E.}\ \bibnamefont
  {Agrapidis}}, \bibinfo {author} {\bibfnamefont {S.-L.}\ \bibnamefont
  {Drechsler}}, \bibinfo {author} {\bibfnamefont {J.}~\bibnamefont {van~den
  Brink}},\ and\ \bibinfo {author} {\bibfnamefont {S.}~\bibnamefont
  {Nishimoto}},\ }\bibfield  {title} {\bibinfo {title} {{Coexistence of
  valence-bond formation and topological order in the Frustrated Ferromagnetic
  $J_1$-$J_2$ Chain}},\ }\href {https://doi.org/10.21468/SciPostPhys.6.2.019}
  {\bibfield  {journal} {\bibinfo  {journal} {SciPost Phys.}\ }\textbf
  {\bibinfo {volume} {6}},\ \bibinfo {pages} {19} (\bibinfo {year}
  {2019})}\BibitemShut {NoStop}%
\bibitem [{\citenamefont {Oshikawa}(1992)}]{Oshikawa92}%
  \BibitemOpen
  \bibfield  {author} {\bibinfo {author} {\bibfnamefont {M.}~\bibnamefont
  {Oshikawa}},\ }\bibfield  {title} {\bibinfo {title} {Hidden
  $\mathrm{Z}_{2}\ast\mathrm{Z}_{2}$ symmetry in quantum spin chains with
  arbitrary integer spin},\ }\href {https://doi.org/10.1088/0953-8984/4/36/019}
  {\bibfield  {journal} {\bibinfo  {journal} {J. Phys.: Condens. Matter}\
  }\textbf {\bibinfo {volume} {4}},\ \bibinfo {pages} {7469} (\bibinfo {year}
  {1992})}\BibitemShut {NoStop}%
\bibitem [{\citenamefont {Cardy}(1996)}]{Cardy96}%
  \BibitemOpen
  \bibfield  {author} {\bibinfo {author} {\bibfnamefont {J.}~\bibnamefont
  {Cardy}},\ }\href {https://doi.org/10.1017/CBO9781316036440} {\emph {\bibinfo
  {title} {Scaling and Renormalization in Statistical Physics}}},\ Cambridge
  Lecture Notes in Physics\ (\bibinfo  {publisher} {Cambridge University
  Press},\ \bibinfo {year} {1996})\BibitemShut {NoStop}%
\bibitem [{\citenamefont {Affleck}\ and\ \citenamefont
  {Ludwig}(1991)}]{Affleck91}%
  \BibitemOpen
  \bibfield  {author} {\bibinfo {author} {\bibfnamefont {I.}~\bibnamefont
  {Affleck}}\ and\ \bibinfo {author} {\bibfnamefont {A.~W.~W.}\ \bibnamefont
  {Ludwig}},\ }\bibfield  {title} {\bibinfo {title} {Universal noninteger
  ``ground-state degeneracy'' in critical quantum systems},\ }\href
  {https://doi.org/10.1103/PhysRevLett.67.161} {\bibfield  {journal} {\bibinfo
  {journal} {Phys. Rev. Lett.}\ }\textbf {\bibinfo {volume} {67}},\ \bibinfo
  {pages} {161} (\bibinfo {year} {1991})}\BibitemShut {NoStop}%
\bibitem [{\citenamefont {Holzhey}\ \emph {et~al.}(1994)\citenamefont
  {Holzhey}, \citenamefont {Larsen},\ and\ \citenamefont
  {Wilczek}}]{Holzhey94}%
  \BibitemOpen
  \bibfield  {author} {\bibinfo {author} {\bibfnamefont {C.}~\bibnamefont
  {Holzhey}}, \bibinfo {author} {\bibfnamefont {F.}~\bibnamefont {Larsen}},\
  and\ \bibinfo {author} {\bibfnamefont {F.}~\bibnamefont {Wilczek}},\
  }\bibfield  {title} {\bibinfo {title} {Geometric and renormalized entropy in
  conformal field theory},\ }\href
  {https://doi.org/https://doi.org/10.1016/0550-3213(94)90402-2} {\bibfield
  {journal} {\bibinfo  {journal} {Nuclear Physics B}\ }\textbf {\bibinfo
  {volume} {424}},\ \bibinfo {pages} {443 } (\bibinfo {year}
  {1994})}\BibitemShut {NoStop}%
\bibitem [{\citenamefont {Calabrese}\ and\ \citenamefont
  {Cardy}(2004)}]{Calabrese04}%
  \BibitemOpen
  \bibfield  {author} {\bibinfo {author} {\bibfnamefont {P.}~\bibnamefont
  {Calabrese}}\ and\ \bibinfo {author} {\bibfnamefont {J.}~\bibnamefont
  {Cardy}},\ }\bibfield  {title} {\bibinfo {title} {Entanglement entropy and
  quantum field theory},\ }\href
  {https://doi.org/10.1088/1742-5468/2004/06/p06002} {\bibfield  {journal}
  {\bibinfo  {journal} {J. Stat. Mech.}\ }\textbf {\bibinfo {volume} {2004}},\
  \bibinfo {pages} {P06002} (\bibinfo {year} {2004})}\BibitemShut {NoStop}%
\bibitem [{\citenamefont {Laflorencie}\ \emph {et~al.}(2006)\citenamefont
  {Laflorencie}, \citenamefont {S\o{}rensen}, \citenamefont {Chang},\ and\
  \citenamefont {Affleck}}]{Laflorencie06}%
  \BibitemOpen
  \bibfield  {author} {\bibinfo {author} {\bibfnamefont {N.}~\bibnamefont
  {Laflorencie}}, \bibinfo {author} {\bibfnamefont {E.~S.}\ \bibnamefont
  {S\o{}rensen}}, \bibinfo {author} {\bibfnamefont {M.-S.}\ \bibnamefont
  {Chang}},\ and\ \bibinfo {author} {\bibfnamefont {I.}~\bibnamefont
  {Affleck}},\ }\bibfield  {title} {\bibinfo {title} {Boundary effects in the
  critical scaling of entanglement entropy in 1$\mathrm{D}$ systems},\ }\href
  {https://doi.org/10.1103/PhysRevLett.96.100603} {\bibfield  {journal}
  {\bibinfo  {journal} {Phys. Rev. Lett.}\ }\textbf {\bibinfo {volume} {96}},\
  \bibinfo {pages} {100603} (\bibinfo {year} {2006})}\BibitemShut {NoStop}%
\bibitem [{\citenamefont {Legeza}\ \emph {et~al.}(2007)\citenamefont {Legeza},
  \citenamefont {S\'olyom}, \citenamefont {Tincani},\ and\ \citenamefont
  {Noack}}]{Legeza07}%
  \BibitemOpen
  \bibfield  {author} {\bibinfo {author} {\bibfnamefont {O.}~\bibnamefont
  {Legeza}}, \bibinfo {author} {\bibfnamefont {J.}~\bibnamefont {S\'olyom}},
  \bibinfo {author} {\bibfnamefont {L.}~\bibnamefont {Tincani}},\ and\ \bibinfo
  {author} {\bibfnamefont {R.~M.}\ \bibnamefont {Noack}},\ }\bibfield  {title}
  {\bibinfo {title} {Entropic analysis of quantum phase transitions from
  uniform to spatially inhomogeneous phases},\ }\href
  {https://doi.org/10.1103/PhysRevLett.99.087203} {\bibfield  {journal}
  {\bibinfo  {journal} {Phys. Rev. Lett.}\ }\textbf {\bibinfo {volume} {99}},\
  \bibinfo {pages} {087203} (\bibinfo {year} {2007})}\BibitemShut {NoStop}%
\bibitem [{\citenamefont {Nishimoto}(2011)}]{cc}%
  \BibitemOpen
  \bibfield  {author} {\bibinfo {author} {\bibfnamefont {S.}~\bibnamefont
  {Nishimoto}},\ }\bibfield  {title} {\bibinfo {title}
  {Tomonaga-luttinger-liquid criticality: Numerical entanglement entropy
  approach},\ }\href {https://doi.org/10.1103/PhysRevB.84.195108} {\bibfield
  {journal} {\bibinfo  {journal} {Phys. Rev. B}\ }\textbf {\bibinfo {volume}
  {84}},\ \bibinfo {pages} {195108} (\bibinfo {year} {2011})}\BibitemShut
  {NoStop}%
\bibitem [{\citenamefont {Totsuka}\ and\ \citenamefont
  {Suzuki}(1995)}]{Totsuka95}%
  \BibitemOpen
  \bibfield  {author} {\bibinfo {author} {\bibfnamefont {K.}~\bibnamefont
  {Totsuka}}\ and\ \bibinfo {author} {\bibfnamefont {M.}~\bibnamefont
  {Suzuki}},\ }\bibfield  {title} {\bibinfo {title} {Matrix formalism for the
  {VBS}-type models and hidden order},\ }\href
  {https://doi.org/10.1088/0953-8984/7/8/012} {\bibfield  {journal} {\bibinfo
  {journal} {J. Phys.: Condens. Matter}\ }\textbf {\bibinfo {volume} {7}},\
  \bibinfo {pages} {1639} (\bibinfo {year} {1995})}\BibitemShut {NoStop}%
\bibitem [{\citenamefont {Nakamura}\ and\ \citenamefont
  {Todo}(2002)}]{Nakamura02}%
  \BibitemOpen
  \bibfield  {author} {\bibinfo {author} {\bibfnamefont {M.}~\bibnamefont
  {Nakamura}}\ and\ \bibinfo {author} {\bibfnamefont {S.}~\bibnamefont
  {Todo}},\ }\bibfield  {title} {\bibinfo {title} {Order parameter to
  characterize valence-bond-solid states in quantum spin chains},\ }\href
  {https://doi.org/10.1103/PhysRevLett.89.077204} {\bibfield  {journal}
  {\bibinfo  {journal} {Phys. Rev. Lett.}\ }\textbf {\bibinfo {volume} {89}},\
  \bibinfo {pages} {077204} (\bibinfo {year} {2002})}\BibitemShut {NoStop}%
\bibitem [{\citenamefont {Miyakoshi}\ \emph {et~al.}(2016)\citenamefont
  {Miyakoshi}, \citenamefont {Nishimoto},\ and\ \citenamefont
  {Ohta}}]{Miyakoshi16}%
  \BibitemOpen
  \bibfield  {author} {\bibinfo {author} {\bibfnamefont {S.}~\bibnamefont
  {Miyakoshi}}, \bibinfo {author} {\bibfnamefont {S.}~\bibnamefont
  {Nishimoto}},\ and\ \bibinfo {author} {\bibfnamefont {Y.}~\bibnamefont
  {Ohta}},\ }\bibfield  {title} {\bibinfo {title} {Entanglement properties of
  the haldane phases: A finite system-size approach},\ }\href
  {https://doi.org/10.1103/PhysRevB.94.235155} {\bibfield  {journal} {\bibinfo
  {journal} {Phys. Rev. B}\ }\textbf {\bibinfo {volume} {94}},\ \bibinfo
  {pages} {235155} (\bibinfo {year} {2016})}\BibitemShut {NoStop}%
\end{thebibliography}%

\end{document}